\documentstyle[preprint,eqsecnum,prd,aps,epsf]{revtex}   
\def\pt{\mbox{$p_T$}}
\def\mpt{p_T}

\newcommand{\eft}{\mbox{$E_{\rm{FRONT}}/E_{\rm{TOTAL}}$}}

\newcommand{\pim}{\mbox{$\pi^{-}$}}
\newcommand{\pip}{\mbox{$\pi^{+}$}}
\newcommand{\kpl}{\mbox{$K^{+}$}}
\newcommand{\piz}{\mbox{$\pi^{0}$}}
\newcommand{\epem}{\mbox{$e^{+}e^{-}$}}
\newcommand{\ycm}{\mbox{$y_{\rm{cm}}$}}
\def\pp{${\mit pp}$}
\def\pBe{${\mit p}$Be}
\def\pA{${\mit p}{\rm A}$}
\def\kt{$k_T$}

\def\avkt{$\langle k_T \rangle$}

\def\PRETH{{\sc pretrigger hi}}
\def\SLL{{\sc single local lo}}
\def\SLH{{\sc single local hi}}
\def\INT{{\sc interaction}}
\def\BEAM{{\sc beam}}
\def\BEAMONE{{\sc beam1}}

\def\LOCALH{{\sc local hi}}
\def\LOCALL{{\sc local lo}}
\def\LOCAL{{\sc local}}

\def\vctr#1{\hfil\vbox to-2ex{\vss\hbox{#1}\vss}\hfil}%
\begin{document}
%
\draft
\preprint{FERMILAB-Pub-01/021-E}
%
%
%
\title{
Production of \piz\ and $\eta$ mesons at large transverse momenta \\
in \pp\ and \pBe\ interactions at 530 and 800~GeV/$c$
}
%
%
\author{                                                                        
L.~Apanasevich,$^{4}$
J.~Bacigalupi,$^{1}$
W.~Baker,$^{3}$
M.~Begel,$^{9}$
S.~Blusk,$^{8}$
C.~Bromberg,$^{4}$
P.~Chang,$^{5}$
B.~Choudhary,$^{2}$
W.~H.~Chung,$^{8}$
L.~de~Barbaro,$^{9}$
W.~DeSoi,$^{9}$
W.~D\l ugosz,$^{5}$
J.~Dunlea,$^{9}$
E.~Engels,~Jr.,$^{8}$
G.~Fanourakis,$^{9}$
T.~Ferbel,$^{9}$
J.~Ftacnik,$^{9}$
D.~Garelick,$^{5}$
G.~Ginther,$^{9}$
M.~Glaubman,$^{5}$
P.~Gutierrez,$^{6}$
K.~Hartman,$^{7}$
J.~Huston,$^{4}$
C.~Johnstone,$^{3}$
V.~Kapoor,$^{2}$
J.~Kuehler,$^{6}$
C.~Lirakis,$^{5}$
F.~Lobkowicz,$^{9,\ddag}$
P.~Lukens,$^{3}$
S.~Mani,$^{1}$
J.~Mansour,$^{9}$
A.~Maul,$^{4}$
R.~Miller,$^{4}$
B.~Y.~Oh,$^{7}$
G.~Osborne,$^{9}$
D.~Pellett,$^{1}$
E.~Prebys,$^{9}$
R.~Roser,$^{9}$
P.~Shepard,$^{8}$
R.~Shivpuri,$^{2}$
D.~Skow,$^{3}$
P.~Slattery,$^{9}$
L.~Sorrell,$^{4}$
D.~Striley,$^{5}$
W.~Toothacker,$^{7,\ddag}$
N.~Varelas,$^{9}$
D.~Weerasundara,$^{8}$
J.~J.~Whitmore,$^{7}$
T.~Yasuda,$^{5}$
C.~Yosef,$^{4}$
M.~Zieli\'{n}ski,$^{9}$
V.~Zutshi$^{2}$
\\
\centerline{(Fermilab E706 Collaboration)}
{~}\\
}                                                                               
\address{                                                                       
\centerline{$^{1}$University of California-Davis, Davis, California 95616}
\centerline{$^{2}$University of Delhi, Delhi, India 110007}
\centerline{$^{3}$Fermi National Accelerator Laboratory, Batavia,              
                   Illinois 60510}                                              
\centerline{$^{4}$Michigan State University, East Lansing, Michigan 48824}     
\centerline{$^{5}$Northeastern University, Boston, Massachusetts  02115}
\centerline{$^{6}$University of Oklahoma, Norman, Oklahoma  73019}
\centerline{$^{7}$Pennsylvania State University, University Park, 
		   Pennsylvania 16802}
\centerline{$^{8}$University of Pittsburgh, Pittsburgh, Pennsylvania 15260}
\centerline{$^{9}$University of Rochester, Rochester, New York 14627}          
\centerline{$^{\ddag}$Deceased}
}                                                                               

\date{\today}
\maketitle
\begin{abstract}
We present results on the production of high transverse momentum
\piz\ and $\eta$ mesons in \pp\ and \pBe\ interactions at 530 and 800~GeV/$c$. 
The data span the kinematic ranges: ${ 1 < \pt\ < 10 \,{\rm GeV}/c}$ 
in transverse momentum and 1.5 units in rapidity. 
The inclusive \piz\  cross sections
are compared with next-to-leading order QCD calculations and to
expectations based on a phenomenological parton-$k_T$ model.
\end{abstract}
\pacs{PACS numbers: 12.38.Qk, 13.85.Ni}
\narrowtext
\section{INTRODUCTION}
The study of inclusive single-hadron production at large transverse 
momentum (\pt) has been a useful probe in the development of
perturbative quantum chromodynamics (PQCD)~\cite{geist,mccubbin}.  
Early in the evolution of the parton model, a departure 
from an exponential dependence of particle production 
at lower \pt\ was interpreted in terms of the onset of 
interactions between pointlike constituents (partons) contained in hadrons.
Large \pt\ is a regime where 
perturbative methods have been applied to QCD to provide quantitative 
comparisons with data.  Such comparisons yield information on the validity
of the PQCD description, and on parton 
distribution functions of hadrons and fragmentation functions of partons.  

This paper reports high-precision measurements of the production of
\piz\ and $\eta$ mesons with large \pt.  The \piz\ production cross
sections are compared with next-to-leading order (NLO) PQCD
calculations~\cite{aversa}.  As illustrated in a previous publication
\cite{prl}, our data (for both inclusive \piz\ and direct-photon
production) are not described satisfactorily by the available NLO PQCD
calculations, using standard choices of parameters.  Similar
discrepancies have been observed \cite{prd} between conventional PQCD
calculations and other measurements of high-\pt\ \piz\ and
direct-photon cross sections (see also \cite{Huston,patrick,frenchpiz}).
The origin of these discrepancies may be attributed to the effects of
initial-state soft-gluon radiation.  Such radiation generates
transverse components of initial-state parton momenta, referred to
below as ${k}_{T}$~\cite{ktnote}.  Evidence of significant \kt\ in
various processes, and a phenomenological model for incorporating its
effect on calculated high-\pt\ cross sections, have been extensively
discussed in Refs.~\cite{prl,prd}; recent studies of the
photoproduction of direct photons at HERA may provide additional
insights \cite{ZEUS-dp,ZEUS-kt,fontannaz-c21,fontannaz-c22,zembrzuski}.
The inadequacy of NLO PQCD in describing \kt-sensitive distributions has 
been discussed in Ref.~\cite{owens-dihad}.  In this paper, we follow the 
phenomenological prescription of Ref.~\cite{prd} when comparing 
calculations with our \piz\ data.  We also present cross sections 
for $\eta$ meson production at large-\pt. As might have been expected from 
earlier measurements (see, $e.g.$, \cite{bigpaper}), $\eta$ production, 
relative to \piz\ production, shows little dependence on \pt\ or on center of
mass rapidity (\ycm).

\section{THE EXPERIMENTAL SETUP}

\begin{figure}
\epsfxsize=6.5truein
\vglue1pt
\vskip1cm
\vglue1pt
\epsfbox[0 72 612 720]{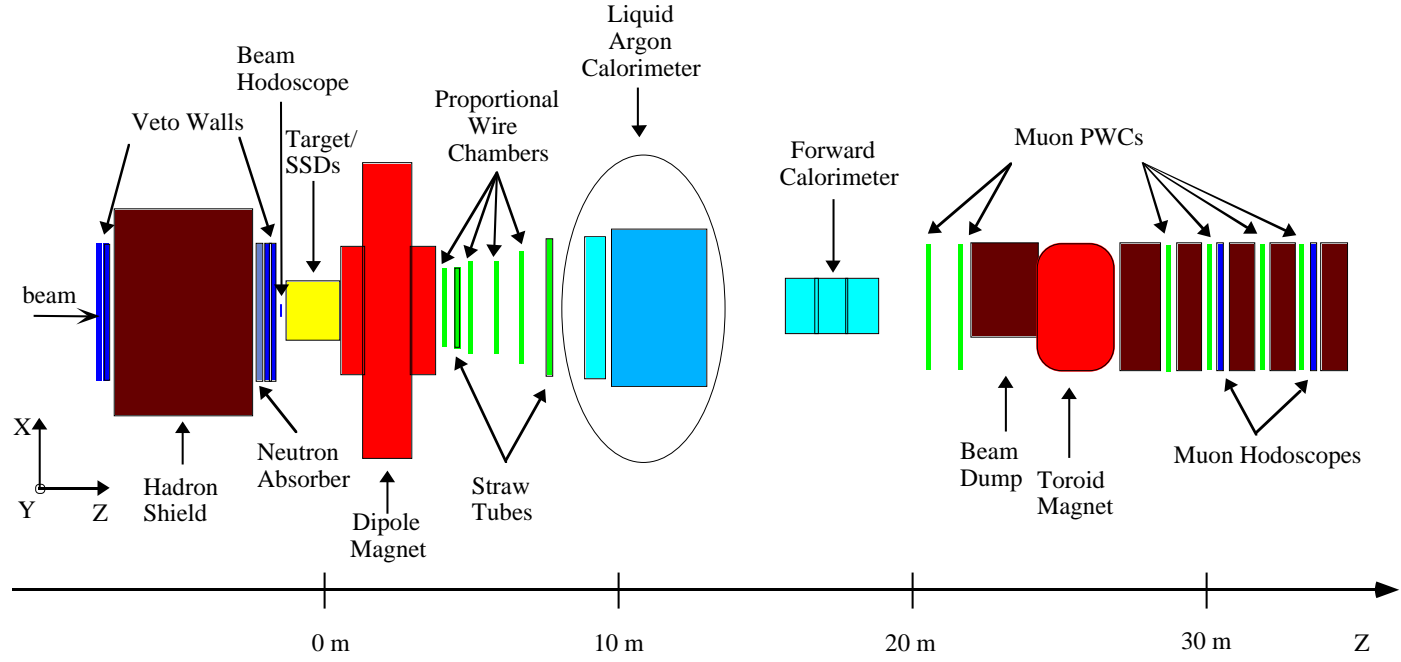}
\vglue1pt
\vskip1cm
\vglue1pt
\caption{
Plan view of the Fermilab Meson West spectrometer, as configured for the 
1991-92 fixed target run.} 
\label{fig:layout}
\end{figure}

The E706 experiment at Fermilab was designed to measure direct-photon
production at high-\pt, and to investigate the structure of events
containing direct photons. The data collection phase of the experiment
spanned three fixed-target running periods, and included a relatively
low statistics commissioning run in 
1987-88~\cite{bigpaper,pi088,dp88,jets88}, and primary data runs in 1990 
and 1991-92. The results presented here are from data recorded during the
1991-92 run. The E706 apparatus, operated in tandem with the E672
muon spectrometer, constituted the Meson West Spectrometer, displayed 
schematically in Fig.~\ref{fig:layout}.
The experiment used a right-handed Cartesian coordinate system, with 
the $Z$-axis pointed in the nominal beam direction, and the $Y$-axis pointed 
upward. The principal elements of the Meson West Spectrometer are discussed 
below. A more detailed description of the triggering methodology appears in 
Ref.~\cite{E706-charm}.

\subsection{Beamline and target}

The Meson West beamline was capable of transporting either 800~GeV/$c$
primary protons from the Fermilab Tevatron or secondary beams of
either polarity. We report here on results from studies
using 800~GeV/$c$ primary protons and a 530~GeV/$c$ positive secondary
beam.  The beamline was instrumented with a differential
\v{C}erenkov counter~\cite{striley,kourbanis} to identify incident
pions, kaons, and protons in secondary beams.  This helium-filled
counter was 43.4~m long and was located $\approx$100~m upstream of the
experiment's target. Using the \v{C}erenkov counter, the proton 
fraction at 530~GeV/$c$ was determined to be 97\%~\cite{striley}.

A 4.7~m long stack of steel surrounding the beam pipe was 
placed between the last beamline magnet and the target box 
(see Fig.~\ref{fig:layout}) to
absorb hadrons. A water tank was placed at the downstream end of this 
hadron shield to absorb low-energy neutrons.  During the 1991-92 run, 
two walls of scintillation counters were located both upstream and 
downstream of the hadron shield to identify penetrating muons.

The target region during the 1991-92 run consisted of two 0.8~mm thick
copper disks of 2.5~cm diameter, located immediately upstream of a
liquid hydrogen target, followed by a 2.54~cm long beryllium cylinder
of diameter 2.54~cm.  The hydrogen target consisted of a 15~cm long
mylar flask, supported in an evacuated volume, with beryllium windows
at each end (2.5~mm thickness upstream and 2.8~mm thickness
downstream) \cite{h2target}.

\subsection{Charged particle tracking}

\begin{figure}
\epsfxsize=5.5truein
\centerline{\epsffile{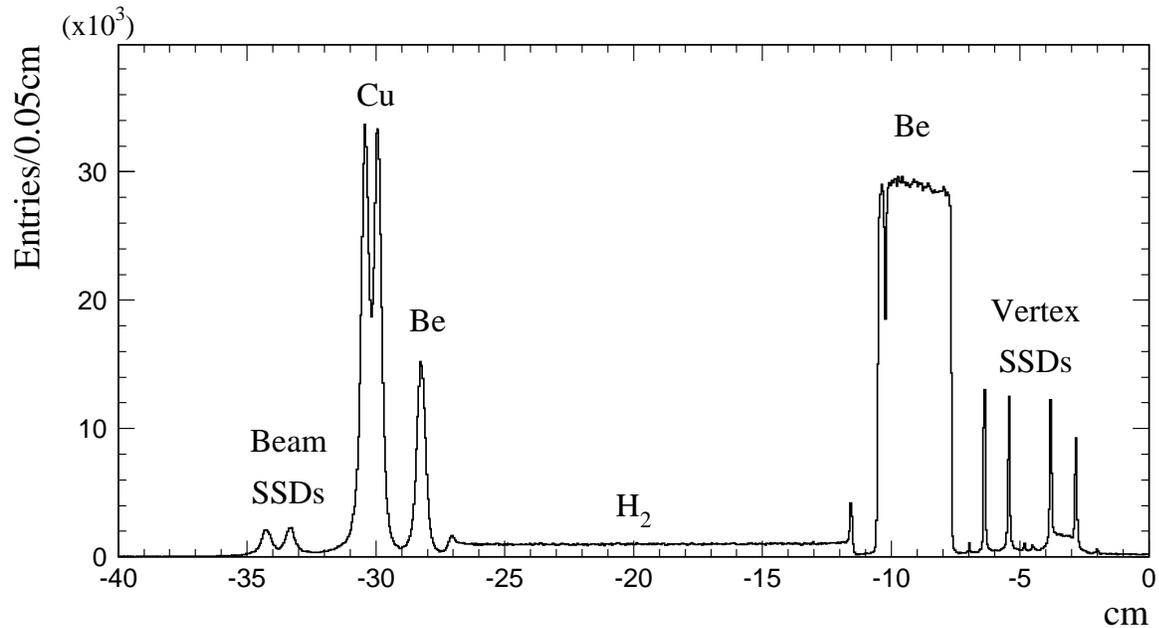}}
\caption{
Distribution of the reconstructed vertex position (along $Z$) for the 
combined 530 and 800~GeV/$c$ data. The clearly-resolved individual 
target elements are, from the left: two copper foils, the liquid hydrogen 
target (a mylar flask enclosed between two thin Be windows), followed by 
the main beryllium target. The two silicon strip detectors (SSDs) 
immediately upstream and four downstream of the target can also be 
easily distinguished.}
\label{fig:target}
\end{figure}

The spectrometer employed a charged particle tracking system consisting of 
silicon strip detectors (SSDs)~\cite{ssd2}, a dipole analysis magnet, 
proportional wire chambers (PWCs), and straw tube drift chambers 
(STDCs)~\cite{straws}. The SSD system consisted of sixteen planes of
silicon wafers, arranged in eight modules. Each module contained 
two SSD planes, one providing $X$-view information and the other
instrumenting the $Y$-view. Six 3$\times$3~cm$^2$ SSD planes were located 
upstream of the target and used to reconstruct beam tracks.  Two 
hybrid 5$\times $5~cm$^2$ SSD planes (25~$\mu$m pitch strips in the 
central 1~cm, 50~$\mu$m beyond) were located $\approx$2~cm downstream of the 
Be target \cite{ssdnote}. These were followed by eight 5$\times $5~cm$^2$ 
SSD planes of 50~$\mu$m pitch.  The SSDs were instrumented to cover a solid 
angle of $\pm 125$~mr. Figure~\ref{fig:target} displays the distribution of 
reconstructed vertices from a representative sample of 1991-92 data,
showing clear separation of the different target elements.

The analysis dipole magnet imparted a transverse momentum impulse of
$\approx$$450~{\rm MeV}/c$ (in the horizontal plane) to singly-charged
particles. Downstream track segments were measured by means of four
stations of four views ($XYUV$) of 2.54~mm pitch PWCs and two stations
of eight (4$X$4$Y$) layers of STDCs with tube diameters 1.03~cm
(upstream station) and 1.59~cm (downstream station). The STDC stations
were installed prior to the 1990 fixed-target run, and improved the
angular resolution of the downstream tracking system (to 
$\approx$0.06~mrad) to make it comparable to that of the upstream system.

\subsection{Calorimetry}

\begin{figure}
\epsfxsize=5.5truein
\vglue1pt
\vskip3cm
\vglue1pt
\hglue1pt
\hskip2cm
\hglue1pt
\epsfbox[0 72 612 720]{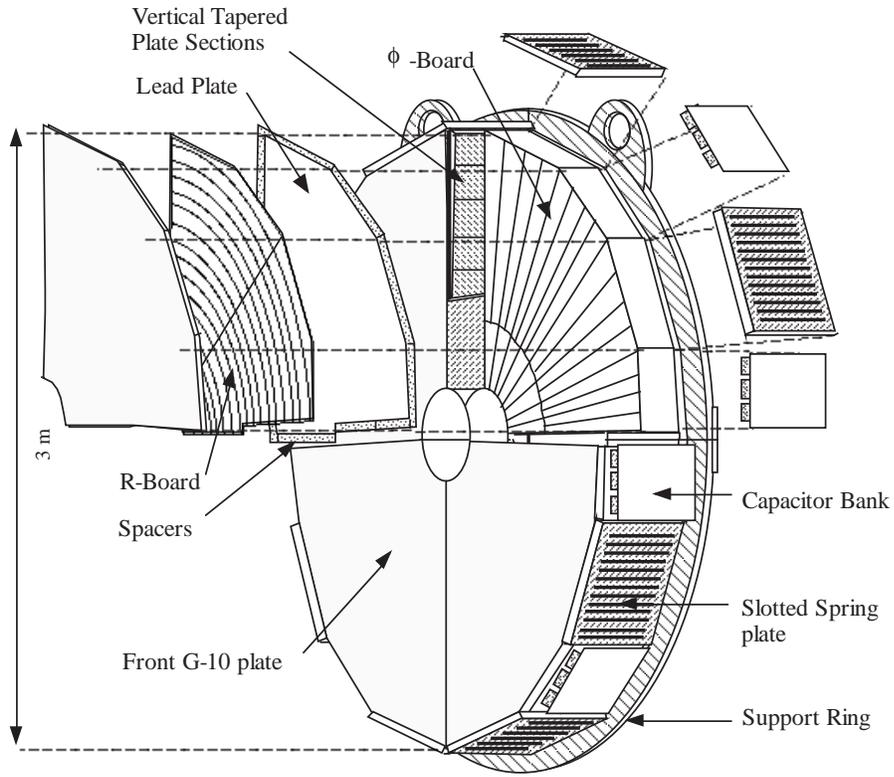}
\vskip-2cm
\caption{
The physical layout of the liquid argon electromagnetic calorimeter (EMLAC).}
\label{fig:emlac}
\end{figure}

The central element of the E706 apparatus was the finely-segmented
liquid argon electromagnetic calorimeter (EMLAC) used to detect and
measure electromagnetic showers~\cite{lac,E706-calibration}. The
EMLAC had a cylindrical geometry with an inner radius of 20~cm and an
outer radius of 160~cm.  It was divided into four mechanically
independent quadrants, which were further subdivided electronically to
create octants.  The calorimeter had 33~longitudinal cells read out in
two sections: an 11 cell front section ($\approx$8.5~radiation
lengths) and a 22 cell back section ($\approx$18~radiation lengths).
This front/back split was used for measuring the direction of
incidence of showering particles, for discriminating between
electromagnetic and hadronic showers and for resolving closely
separated electromagnetic showers. The longitudinal cells consisted of
2~mm thick lead cathodes (the first cathode was constructed of
aluminum), double-sided copper-clad G-10 radial ($R$) anode boards,
followed by 2~mm thick lead cathodes and double-sided copper-clad G-10
azimuthal ($\Phi$) anode boards. There were 2.5~mm argon gaps between
each of these layers in a cell. The physical layout is illustrated in
Fig.~\ref{fig:emlac}.

The copper-cladding on the anode boards was cut to form strips. Signals from
corresponding strips from all $R$ (or $\Phi$) anode boards in the front
(or back) section were jumpered together.
The copper-cladding on the radial anode boards was cut into concentric strips
centered on the nominal beam axis. The width of the strips on the first
$R$ board was 5.5~mm. The width of the $R$ strips on the following $R$ boards
increased slightly so that the radial geometry was projective relative
to the target, which was located 9~m upstream of the EMLAC.
The azimuthal readout was subdivided at a radius of 40~cm 
into inner and outer segments, with each inner $\Phi$ strip subtending an 
azimuthal angle of $\pi/192$~radians, and outer $\Phi$ strips covering 
$\pi/384$~radians.   
Subdivision of the azimuthal strips in the outer portion of the detector 
improved both the position and energy resolution for showers reconstructed 
in this region.  It also reduced $R$--$\Phi$ correlation ambiguities from
multiple showers in the same octant of the calorimeter.

Data acquisition and trigger-signal processing for the EMLAC was based 
upon the FNAL RABBIT system~\cite{rabbit}.  To achieve tolerable deadtimes,
the zero suppression features of this system were used extensively during the 
experiment's 1987-88 commissioning run.  However, zero suppression limited
our ability to understand the effects of out-of-time photon-induced showers
and the tails of hadron-induced showers, thereby compromising efforts to 
understand the detailed response of the detector. Consequently, FASTBUS 
modules (the ICBM---Intelligent Control and Buffering Module---and the Wolf 
interface~\cite{ICBM,wolf}) were developed by E706 to replace the original, 
CDF-designed, MX readout controllers.  These FASTBUS modules enabled us to 
eliminate zero suppression during the experiment's two primary data runs.

The apparatus also included two other calorimeters: a hadronic calorimeter 
(HALAC) located downstream of the EMLAC in the same cryostat, and a 
steel and scintillator calorimeter (FCAL), positioned further downstream, to 
increase coverage in the very forward region.  The HALAC had 
53~longitudinal cells read out in 2~sections: a 14~cell front section 
($\approx$2 interaction lengths), and a 39~cell back section ($\approx$6 
interaction lengths).  Each cell consisted of read-out planes separated by 
3~mm argon gaps and a 2.5~cm thick steel plate.

The FCAL acceptance covered the beam hole region of the EMLAC. It was split 
into three longitudinally similar sections. Each section was composed of 
alternating layers of 1.9~cm thick steel absorber plates and 4.8~mm thick 
acrylic scintillator sheets.  The distance between steel plates 
was 6.9~mm.  The downstream module contained 32 steel absorber plates 
and 33 scintillator sheets; the other two modules were comprised of 28 steel
absorber plates and 29 scintillator sheets.  Together, the three modules 
constituted $\approx$10.5~interaction lengths of material.
  
\subsection{Muon identification}

The E672 muon spectrometer, consisting of a toroidal magnet, shielding,
scintillators, 
and proportional wire chambers, was deployed immediately downstream of the 
FCAL.  The combined Meson West Spectrometer was triggered on high-mass muon 
pairs in order to investigate the hadroproduction of $J/\psi$, $\psi(2s)$, 
$\chi_c$, and B mesons~\cite{psi90,psi91,chi90,B90}.  E706 and E672 
collected data simultaneously and shared trigger logic, the data acquisition
system, and event reconstruction programs.  Data collected with the dimuon 
trigger were also used for several technical studies---for example, the 
$J/\psi\longrightarrow\mu^+\mu^-$ signal was used to calibrate the momentum 
scale of the tracking system.

\subsection{Triggering}

The E706 trigger selected events yielding high transverse momentum
showers in the EMLAC. The selection process involved four stages: beam
and interaction definitions, pretrigger requirements, and the final
trigger requirements. Beam particles were detected using a hodoscope
consisting of three planes (arranged in $X$, $Y$ and $U$ views) of
scintillator, located $\approx$2~m upstream of the target region.
Each plane contained 12 scintillator strips. The widths of the strips
varied from 1~mm in the central region, to 5~mm along the edges.  The
edges of individual strips in each plane were overlapped to avoid gaps
in acceptance. A \BEAM\ signal was generated by a coincidence of
signals from counters in at least two of the three hodoscope planes.
A \BEAMONE\ signal required that less than two hodoscope planes
detected two or more isolated clusters of hits in coincidence with
\BEAM.  This \BEAMONE\ requirement rejected events with multiple beam
particles incident upon the target. A plane of four scintillation
counters (referred to as beam hole counters) arranged to produce a
0.95~cm diameter central hole was located downstream of the beam
hodoscope and used to reject interactions initiated by particles in
the beam halo.

Two pairs of scintillation counters were mounted on the dipole
analysis magnet, one pair upstream and the other downstream of the
magnet.  Each pair had a central hole that allowed non-interacting
beam particles to pass through undetected. An \INT\ was defined as a
coincidence between signals from at least two of these four
interaction counters.  To minimize potential confusion in the EMLAC
due to out-of-time interactions, a filter was used to reject
interactions that occurred within 60~ns of one other.

For those interactions that satisfied both the \BEAMONE\ and \INT\
definitions, the $p_T$ deposited in various regions of the EMLAC was
evaluated by weighting the energy signals from the fast outputs of the
EMLAC $R$ channel amplifiers by $\approx$$\sin{\theta_i}$, where
$\theta_i$ is the polar angle that the $i^{th}$ strip subtends with
respect to the nominal beam axis.  The \PRETH\ requirement was
satisfied when the $p_T$ detected either in the inner 128~$R$ channels
or the outer $R$ channels of any octant was greater than the threshold
value of $\approx$2~GeV/$c$. A pretrigger signal was issued only if
the signals from a given octant satisfied the pretrigger requirement,
there was no evidence in that octant of substantial noise or
significant \pt\ attributable to an earlier interaction, and there was
no incident beam halo muon detected by the walls of scintillation
counters surrounding the hadron shield upstream of the spectrometer.
The pretrigger signal latched data from the various subsystems until a
final trigger decision was made.

Localized trigger groups were formed for each octant by clustering
the $R$ channels into 32~groups of 8~channels.  Each of the adjacent pairs 
of 8~channel groups (groups 1 and 2, 2 and 3, $\ldots$~, 31 and 32) 
formed a \LOCAL\ group of 16 strips.  If the $p_T$ detected 
in any of these groups of 16 was above a specified high (or low) threshold, 
then a \LOCALH\ (or \LOCALL) signal was generated for that octant.  The 
\SLH\ (\SLL) trigger required a \LOCALH\ (\LOCALL) signal from an octant 
that also satisfied the \PRETH. The \LOCALH\ (\LOCALL) threshold 
was $\approx$3.5~GeV/$c$ ($\approx$2~GeV/$c$). The \SLL\ trigger was 
prescaled by a factor of $\approx$200.

In addition to these high-$p_T$ triggers, prescaled samples of low-bias 
\BEAM, \INT, and pretrigger events were also recorded. The prescale
factors for these triggers were typically set at $15^6$, $15^5$, and $15^3$, 
respectively. These low-bias triggers constituted $\approx$10$\%$ 
of the events recorded.

\section{\piz\ AND $\eta$ ANALYSIS}

The data sample used in this analysis corresponds to an integrated
luminosity of 6.8 (1.2) events/pb for 530~GeV/$c$ \pBe\ (\pp) interactions,
and 6.5 (1.2) events/pb at 800~GeV/$c$. The following subsections describe 
the analysis procedures and methods used to correct the data for losses 
resulting from inefficiencies and selection biases.

\subsection{Event reconstruction}

The charged-track reconstruction algorithm produced track segments
upstream of the dipole magnet using information from the SSDs, and
downstream of the magnet using information from the PWCs and
STDCs. These track segments were projected to the center of the
magnet, and linked to form the final tracks, and the interaction vertex.
The charged-track reconstruction and vertex-finding methodology are 
described in Ref.~\cite{E706-charm}.

The readout in each EMLAC quadrant consisted of four regions: left $R$
and right $R$ (radial strips of each octant in that quadrant), and
inner $\Phi$ and outer $\Phi$ regions (azimuthal strips divided at $R
= 40~{\rm cm}$).  Strip energies from clusters in each region were
fitted to the shape of electromagnetic showers, as determined from
Monte Carlo simulations and isolated-shower data.  These fits were
used to evaluate the positions and energies ($E_R$ and $E_\Phi$) of
the peaks in each region.  Shower positions and energies were obtained
by correlating peaks of approximately the same energy in the $R$ and
$\Phi$ regions within the same half octant (more complex algorithms
were used to handle configurations with overlapping showers in either
the $R$ or $\Phi$ regions). Any differences in photon energy as
measured in the $R$ and $\Phi$ views reflect the intrinsic resolution
properties of the calorimeter, and provide a test of the quality of
the Monte Carlo simulations. The EMLAC's longitudinal segmentation
provided discrimination between showers generated by
electromagnetically or hadronically interacting particles. For
individual showers, the ratio of energy reconstructed in the front
section to the sum of energy in the front and back section of the
EMLAC (referred to as \eft) also tested the Monte Carlo simulation of
longitudinal shower development (see the detector simulation section
below).  An expanded discussion of the EMLAC reconstruction procedures
and performance can be found in Ref.~\cite{E706-calibration}.

\subsection{Data sample selection and corrections}

The events contributing to the cross section measurements were
required to have a reconstructed vertex within the fiducial volume of
the Be or H${}_2$ targets \cite{htargnote}. Vertex reconstruction
efficiencies were evaluated for each target using a detailed Monte
Carlo simulation of the spectrometer (described below). These
efficiencies were used to correct for reconstruction losses and for
resolution smearing across target fiducial boundaries. The vertex
reconstruction efficiency was $\approx 1$ for the H${}_2$ and
downstream Be targets and $0.97$ for the upstream Be target.

Both \piz\ and $\eta$ mesons were reconstructed via their
$\gamma\gamma$ decay modes. The photons were required to be within the
fiducial region of the EMLAC, which excluded areas with reduced
sensitivity.  In particular, photons incident upon regions of the
detector near quadrant boundaries (which abutted steel support
plates), the central beam hole, the outer radius of the EMLAC, and
octant boundaries were excluded from consideration. A simple
ray-tracing Monte Carlo program was employed to determine the
correction for the effect of the azimuthal fiducial boundaries
\cite{accnote}.

To reduce backgrounds from hadronic showers, only showers with at
least 20$\%$ of their shower energy deposited in the front part of
EMLAC (\eft $>$ 0.2) were accepted as photon candidates. To correct
the cross sections for this requirement, as well as for other larger
effects including resolution-smearing and reconstruction losses, a
full simulation of the showers in the calorimeter (described below)
was employed.  In addition, $\gamma\gamma$ combinations were
considered as \piz\ or $\eta$ candidates only when the two photons
were detected in the same octant (to simplify subsequent analysis of
the trigger response), and only those combinations with energy asymmetry 
[$A_{\gamma\gamma} \equiv |E _{\gamma_1}-E_{\gamma_2}|/(E_{\gamma_1}+ 
E_{\gamma_2})$] less than 0.75 were considered (to reduce uncertainties 
due to low energy photons).

\begin{figure}
\epsfxsize=5.5truein
\vglue1pt
\vskip3cm
\vglue1pt
\hglue1pt
\hskip2cm
\hglue1pt
\epsfbox[0 72 612 720]{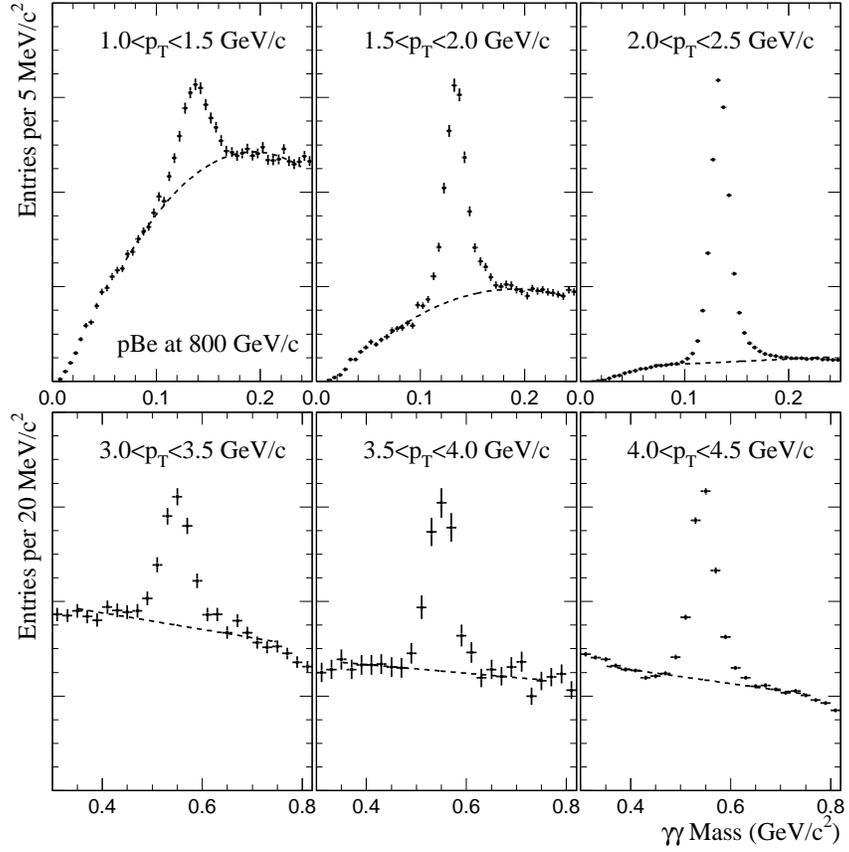}
\caption{
$\gamma\gamma$ mass distributions in the region of the \piz\ (top row)
and the $\eta$ (bottom row) mesons from \pBe\ interactions at 800~GeV/$c$,
for several ranges of $\gamma\gamma$ \pt-values.}
\label{fig:mvpt}
\end{figure}

The $\gamma\gamma$ invariant mass distribution in the \piz\ and $\eta$
regions for photon pairs that satisfied the above requirements is
shown in Fig.~\ref{fig:mvpt} for representative low \pt\ intervals. A
\piz\ candidate was defined as a combination of two photons with
invariant mass, $M_{\gamma \gamma}$ \cite{massnote}, in the range
$100~{\rm MeV}/c^2 < M_{\gamma\gamma} < 180~{\rm MeV}/c^2$.
An $\eta$ candidate was defined as a two-photon combination in the
range $450~{\rm MeV}/c^2 < M_{\gamma\gamma} < 650~{\rm MeV}/c^2$. To
evaluate production cross sections, combinatorial background in the
\piz\ and $\eta$ regions was evaluated as follows. Sideband regions were 
defined to cover an equivalent mass range of the \piz\ and $\eta$ peak
regions (using the same acceptance criteria as for the peak regions).
The \pt\ and rapidity distributions from these side bands were then
subtracted from the corresponding distributions within the \piz\ and
$\eta$ mass ranges to obtain the respective signals. This technique is
appropriate as long as the combinatorial background depends
approximately linearly dependence on $M_{\gamma\gamma}$.  At low \pt\
(below $\approx 2$~GeV/$c$), the shape of the combinatorial background
in the signal regions is not linear, and a more sophisticated fitting
procedure was used to evaluate the background. The $\gamma\gamma$ mass
distributions were fitted to Gaussians for signal plus second-order
polynomials in $M_{\gamma\gamma}$ to represent the background. The
combinatorial background in the peak regions was then defined using
the resultant parameters of the fit, and the signals defined as the
differences between the totals and the fitted backgrounds.

For the cross section measurements, the signals have been corrected
for losses due to the energy asymmetry cut and the branching fractions
\cite{pdg} for the $\gamma\gamma$ decay modes.  The correction for
losses due to the conversion of one or both of the photons into \epem\
pairs was evaluated by projecting each reconstructed photon from the
event vertex to the reconstructed position in the EMLAC.  The number
of radiation lengths of material traversed along the photon path was
calculated on the basis of a detailed description of the detector.
The photon conversion probability was then evaluated and used to
account for losses from conversion.

\subsection{Trigger response}

As mentioned previously, the \SLH\ and \SLL\ trigger decisions were
based upon depositions of \pt\ in the EMLAC within groups of 16
contiguous radial strips. For each \piz\ or $\eta$ candidate, a
probability to satisfy the trigger was defined, based upon energy
deposition in the entire octant, as $P = 1 -\prod(1-p_{\mit i})$, 
where $p_{\mit i}$ is the efficiency of the ${\mit i}^{th}$ trigger
group in the octant containing the candidate \cite{trignote}. The
inverse of this probability was applied as a trigger weight to each
meson candidate. To avoid excessively large trigger weights, meson
candidates with trigger probabilities of $<$ 0.1 were excluded from
further consideration. The correction for losses from this requirement
was determined from Monte Carlo, and absorbed into the reconstruction
efficiency.

The efficiency of the \PRETH\ was determined in a manner similar to
that used for the local triggers \cite{trignote}. Additional details
on the trigger can be found in Refs.~\cite{E706-charm,sorrell,osborne}.

\begin{figure}
\epsfxsize=5.5truein
\centerline{\epsffile{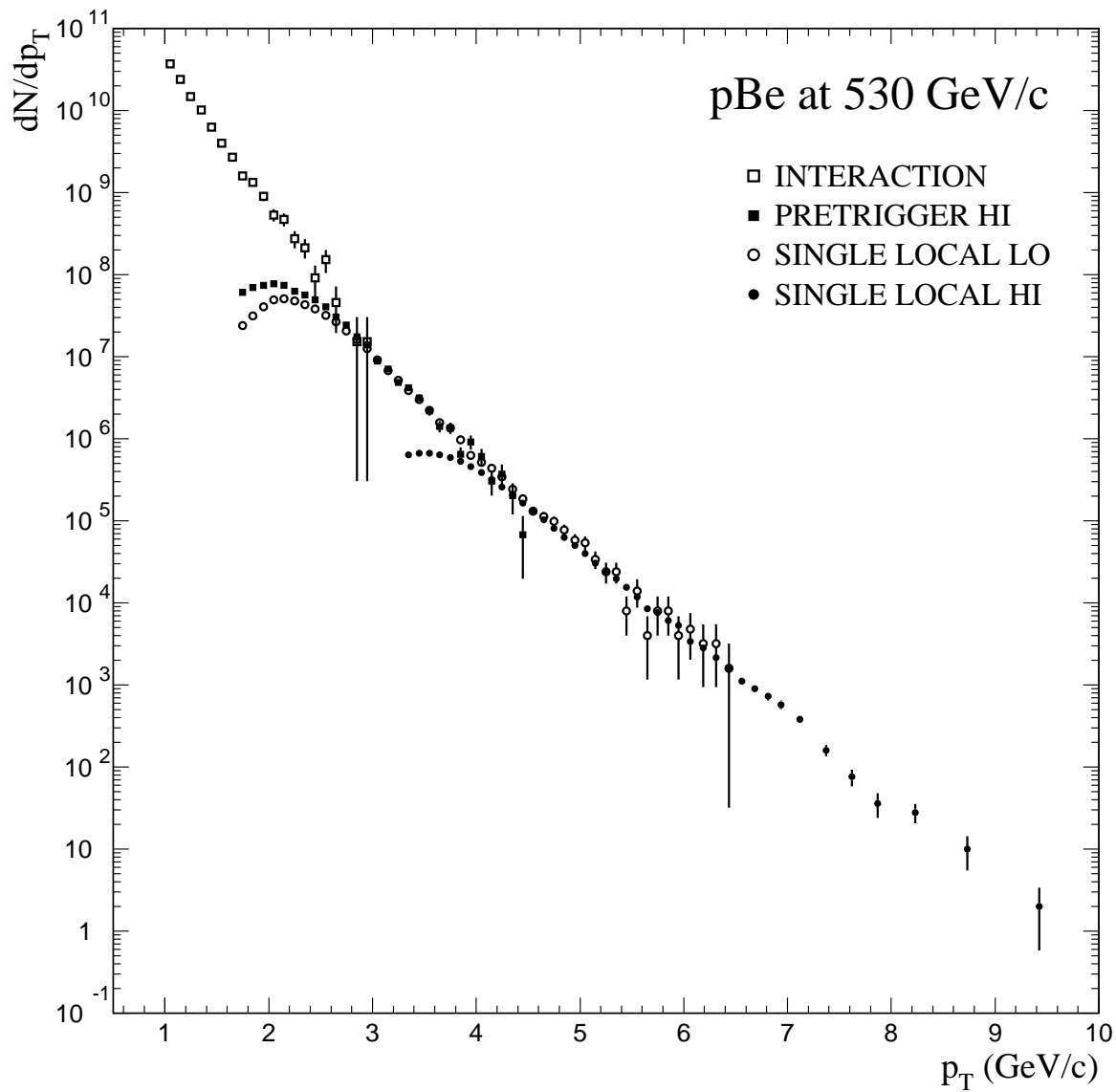}}
\caption{
\pt\ spectra of \piz\ candidates from 530~GeV/$c$ proton interactions on
Be, selected via several different triggers implemented with significantly 
different prescale factors. The data have only been corrected for trigger 
prescale factors.
\label{fig:pixs_full}}
\end{figure}

The cross sections presented in this paper reflect composite
measurements, utilizing a combination of results from the \INT,
\PRETH, \SLL, and \SLH\ triggers. The \pt\ spectra (corrected for only
prescale factors) from a representative sample of these triggers are
shown in Fig.~\ref{fig:pixs_full}. The transition points chosen
between the high and low threshold triggers were determined by
comparing the fully corrected results from each trigger, and were
different for \piz\ and $\eta$ mesons and also depended on rapidity.

\subsection{Rejection of beam halo muons}

Spurious triggers were produced by muons in the beam halo that
radiated energy in the electromagnetic calorimeter in random
coincidence with an interaction in the target. Particularly in the
outer regions of the EMLAC, this energy was recorded as a high-$p_T$
deposition that satisfied the \LOCAL\ trigger requirements.  This
occurred much more frequently in data from the 530~GeV/$c$ secondary 
beam than for the primary 800~GeV/$c$ beam, due to the absence
of an upstream interaction target in the latter case. To reduce this
background, the pretrigger logic relied on signals from the veto walls
of scintillator counters to reject events associated with such muons
in the beam halo.  In the off-line analysis, we employed expanded
requirements on the latched veto-wall signals, the direction of
reconstructed showers \cite{E706-calibration}, the shower shape (halo
muon-induced showers have a different shape than photon or
electron-induced showers), and the total $p_T$ (balance) in the event.
For the latter, we calculated the net \pt\ of the photons and charged
particles which, based upon their initial trajectories, would
intercept the EMLAC in the transverse plane within the $120^\circ$
sector opposite the meson candidate. In interactions which generate a
high-\pt\ meson, the ratio of this ``away-side'' \pt\ ($P_T^{away}$)
to that of the meson \pt\ should be near unity. However, for events
triggered by showers from beam halo muons, $P_T^{away}/ \mpt $ should
be near zero, since the interaction in random coincidence with the
beam halo muon is typically a soft (low \pt) interaction. Candidates
with $P_T^{away}/\mpt < 0.3$ were considered likely to be due to beam
halo muons and were rejected.

\begin{figure}
\epsfxsize=5.5truein
\epsfbox[0 72 612 720]{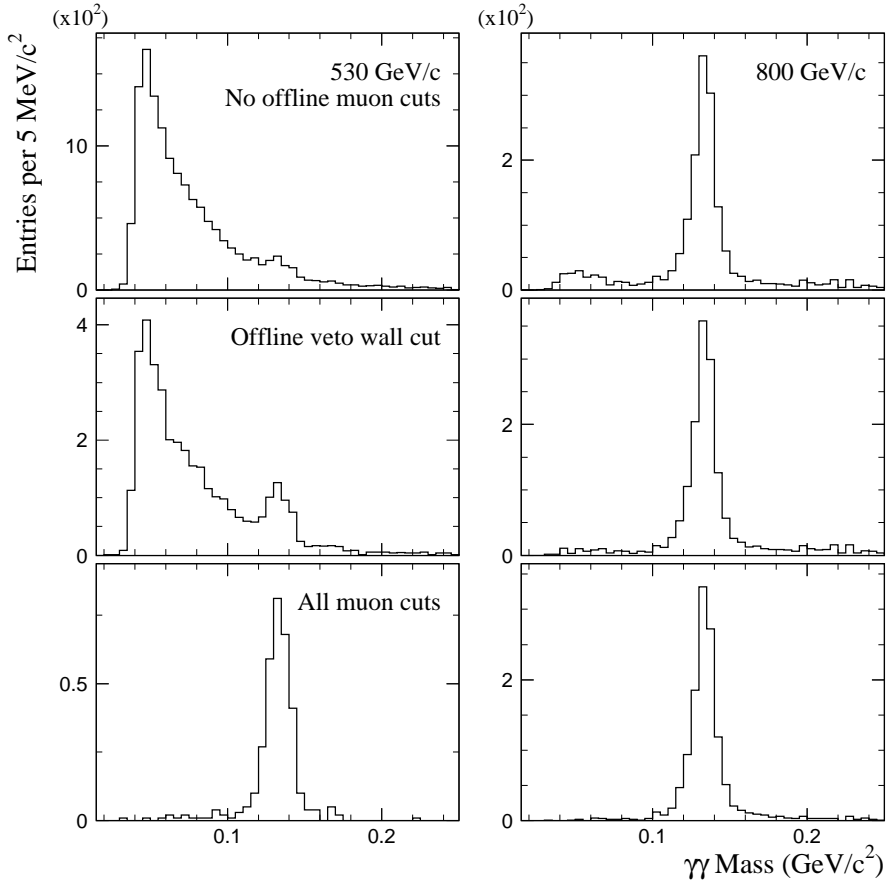}
\caption{
Effect of beam halo muon induced shower discriminators on the $\gamma\gamma$ 
invariant mass distribution in the \piz-mass region, for $\gamma\gamma$  
pairs with $p_T>7$~GeV/$c$ in the 530~GeV/$c$ data (left) and in the 
800~GeV/$c$ data (right).}
\label{fig:muons}
\end{figure}

To illustrate the effect of the above off-line requirements, the
$\gamma\gamma$ invariant mass distribution, both before and after
application of the rejection criteria, is shown in
Fig.~\ref{fig:muons} for the 530 and 800~GeV/$c$ data, for $p_T > 7.0$
GeV/c.  The large muon-induced background at low $\gamma\gamma$ mass
values in the 530~GeV/$c$ data is due to the occasional transverse
splitting of the muon-induced showers into two closely-separated
photon candidates. This happens because the reconstruction software
assumes that the showers originate from the target-region rather than
from the beam halo.  The 800~GeV/$c$ data have very few muon-induced
triggers, and is consequently not affected very strongly by these
rejection criteria.

The impact of these rejection criteria on the physics signal was
checked using more restrictive selection criteria to define a pure
sample of $\gamma\gamma$ pairs. The fraction of signal lost by the
application of each of the muon-rejection requirements determined a
correction to the cross section. The product of the correction factors
for muon rejection corresponds to an increase of $\approx$8$\%$ in the
cross secton at $p_T$ = 4~GeV/$c$, and $\approx$10$\%$ at $p_T = 7$
GeV/$c$, for the 530~GeV/$c$ beam data. (The corrections were smaller
for the 800~GeV/$c$ data because of cleaner beam conditions.)

\subsection{Detector simulation}

The Meson West spectrometer was modeled using a detailed {\sc
geant}~\cite{geant} simulation. Because the full simulation of
electromagnetic showers requires extensive computing time, we
developed a hybrid approach using {\sc geant}-tracking through the
magnetic spectrometer and in the initial stages of the shower
development in the calorimeter.  We used the standard {\sc geant}
algorithms for tracking particles with energies above 10 MeV, below
which we relied on an empirical parameterization for the deposition of
energy in the EMLAC \cite{88mcnote}. This cutoff was selected to be at
the point at which bremsstrahlung still dominates the energy loss in
lead, and led to significant improvement in processing speed.  In
doing this, we took advantage of the steady advances in computational
power of the FNAL UNIX farms~\cite{farm} to reach the desired level of
statistical accuracy.

As inputs to the {\sc geant} simulation, we employed single particle
distributions, reconstructed data events, and two physics event
generators: {\sc herwig}~\cite{herwig56} and {\sc
pythia}~\cite{pythia56}.  For the analyses described herein, we chose
{\sc herwig} as the principal Monte Carlo event generator based on a
better match of particle multiplicities between data and Monte Carlo
using the default parameters. Over 5.5~million {\sc herwig} events
were passed through the {\sc geant} simulation. We weighted the {\sc
herwig} \piz\ and $\eta$ spectra (in $p_T$ and rapidity) to match our
measured results, so that the corrections obtained from the Monte
Carlo were based on the data distributions rather than on the behavior
of the physics generator.

The calibration of the energy response of the EMLAC was based on the
reconstructed masses of \piz\ mesons in the $\gamma\gamma$ decay
mode~\cite{E706-calibration}.  The steeply falling $p_T$ spectrum for
\piz\ production, combined with the calorimeter's resolution, produced
a small offset ($\approx$1$\%$) in the mean reconstructed photon
energies. Using the same calibration procedure in the simulated EMLAC
as in the detector, we corrected this offset and minimized any
potential biases in the calibration.  We also employed the {\sc geant}
Monte Carlo simulation to evaluate the mean correction (as a function
of photon energy) for energy deposited in the material upstream of the
EMLAC.  The impact of detector resolution on the energy scale and on
the \piz\ and $\eta$ production spectra was incorporated in the
overall reconstruction efficiency corrections.

To ensure that the Monte Carlo simulation reproduced the data, a
special preprocessor was used to convert {\sc geant} information into
signals and strip energies as measured in the various detectors, and
to simulate hardware effects, such as channel noise and gain
variations.  The generated Monte Carlo events were then processed
through the same reconstruction software used for the analysis of
data, and thereby provided measures of the inefficiencies and biases
of the reconstruction algorithms.

\begin{figure}
\epsfxsize=5.5truein
\epsfbox[0 72 612 720]{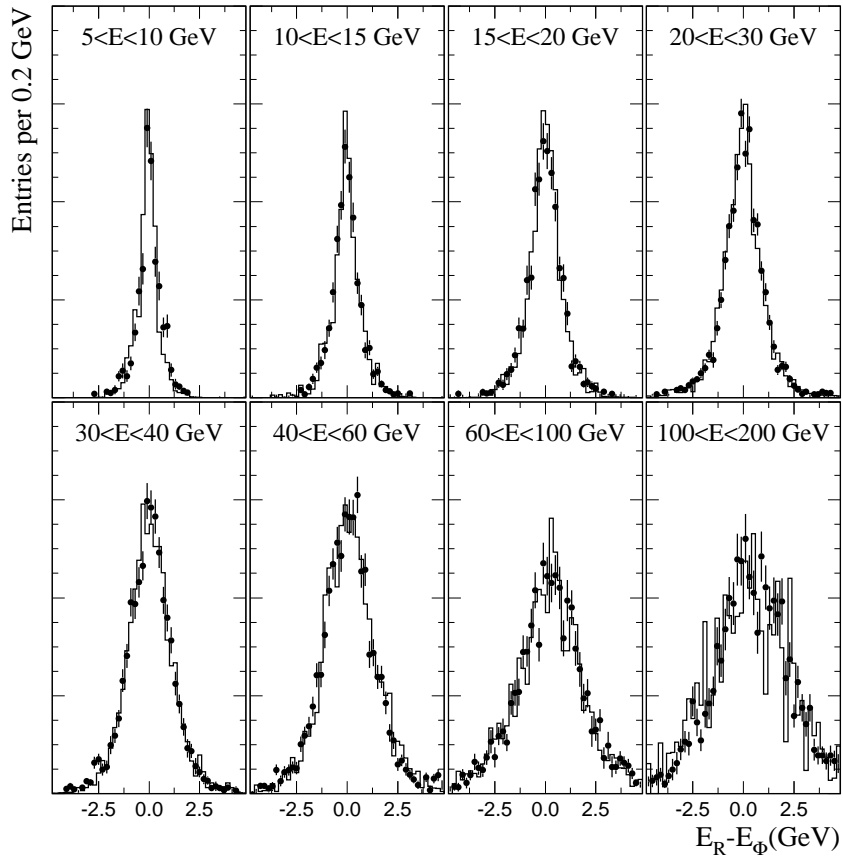}
\caption{
Comparisons of distributions in photon $E_R-E_\Phi$, for eight photon-energy 
bins, between data (histogram) and the Monte Carlo simulation (points) for 
the 530~GeV/$c$ sample. The photons are from \piz\ candidates with 
$p_T>3.5$~GeV/$c$.}
\label{fig:eref}
\end{figure}

\begin{figure}
\epsfxsize=5.5truein
\epsfbox[0 72 612 720]{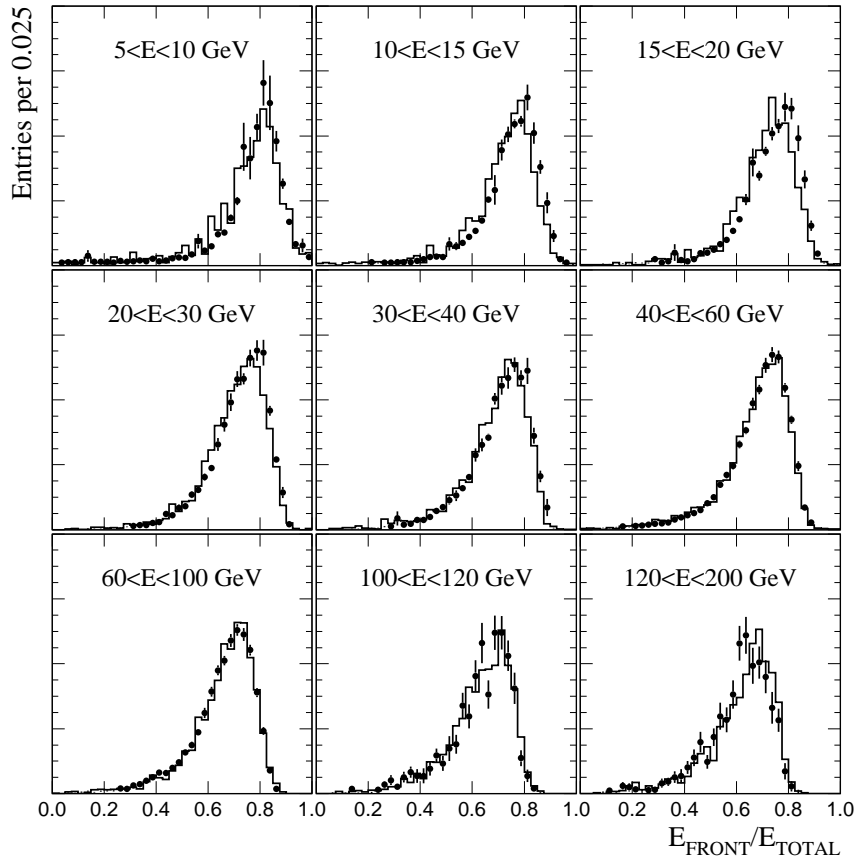}
\caption{
Comparison of the photon \eft\ distributions, in nine photon-energy bins, 
between data (histogram) and the Monte Carlo simulation (points), for 
the 800~GeV/$c$ sample. The photons originated from decays of \piz\ 
candidates with $p_T>3.5$~GeV/$c$.}
\label{fig:efet}
\end{figure}

\begin{figure}
\epsfxsize=5.5truein
\epsfbox[0 72 612 720]{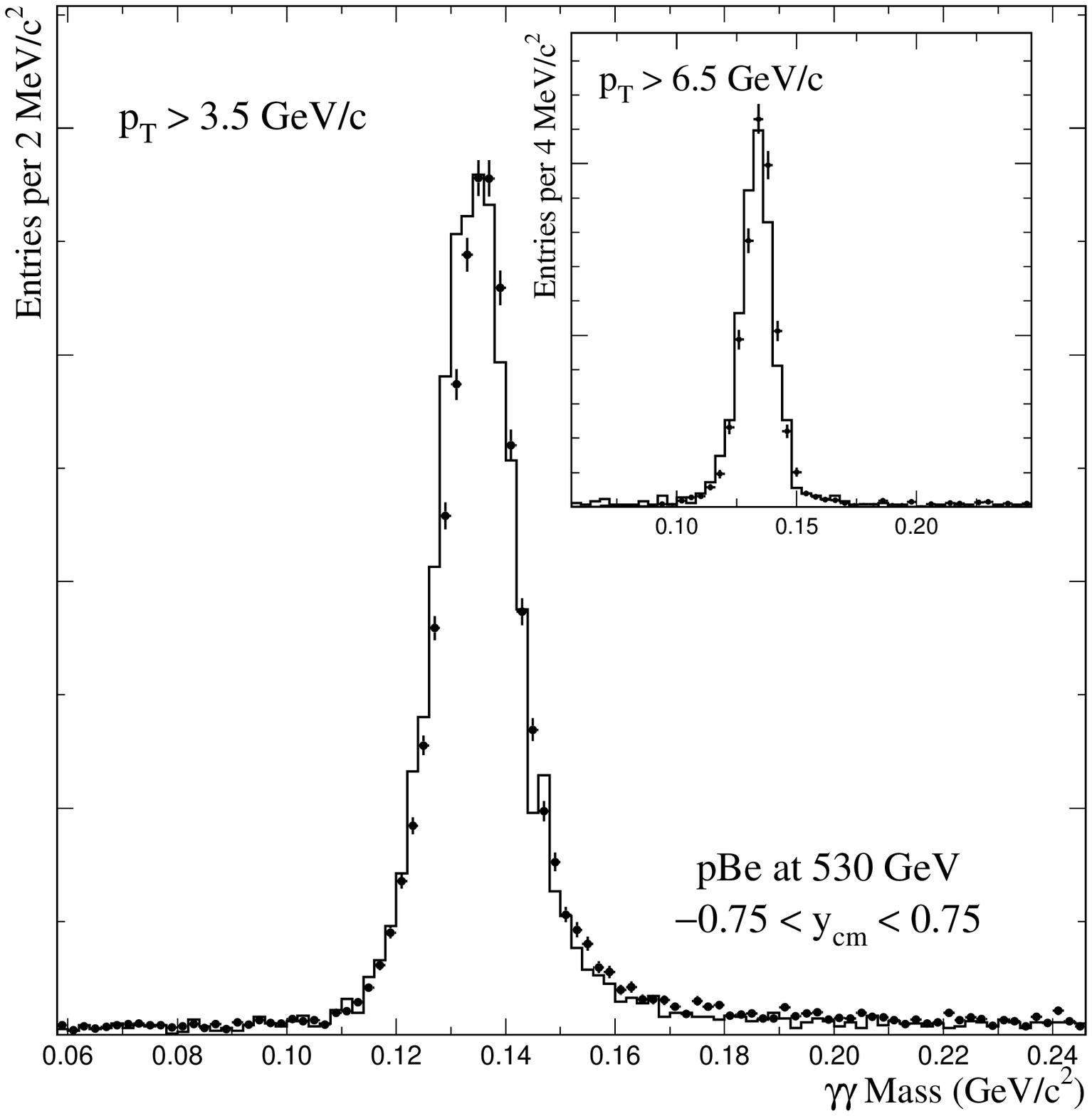}
\caption{
$\gamma\gamma$ mass distributions in the \piz\ signal region in the 530 
GeV/$c$ data (histogram) compared to Monte Carlo simulations (points) for two 
requirements on minimum $p_T$.}
\label{fig:comp_mass_pi0}
\end{figure}

\begin{figure}
\epsfxsize=5.5truein
\epsfbox[0 72 612 720]{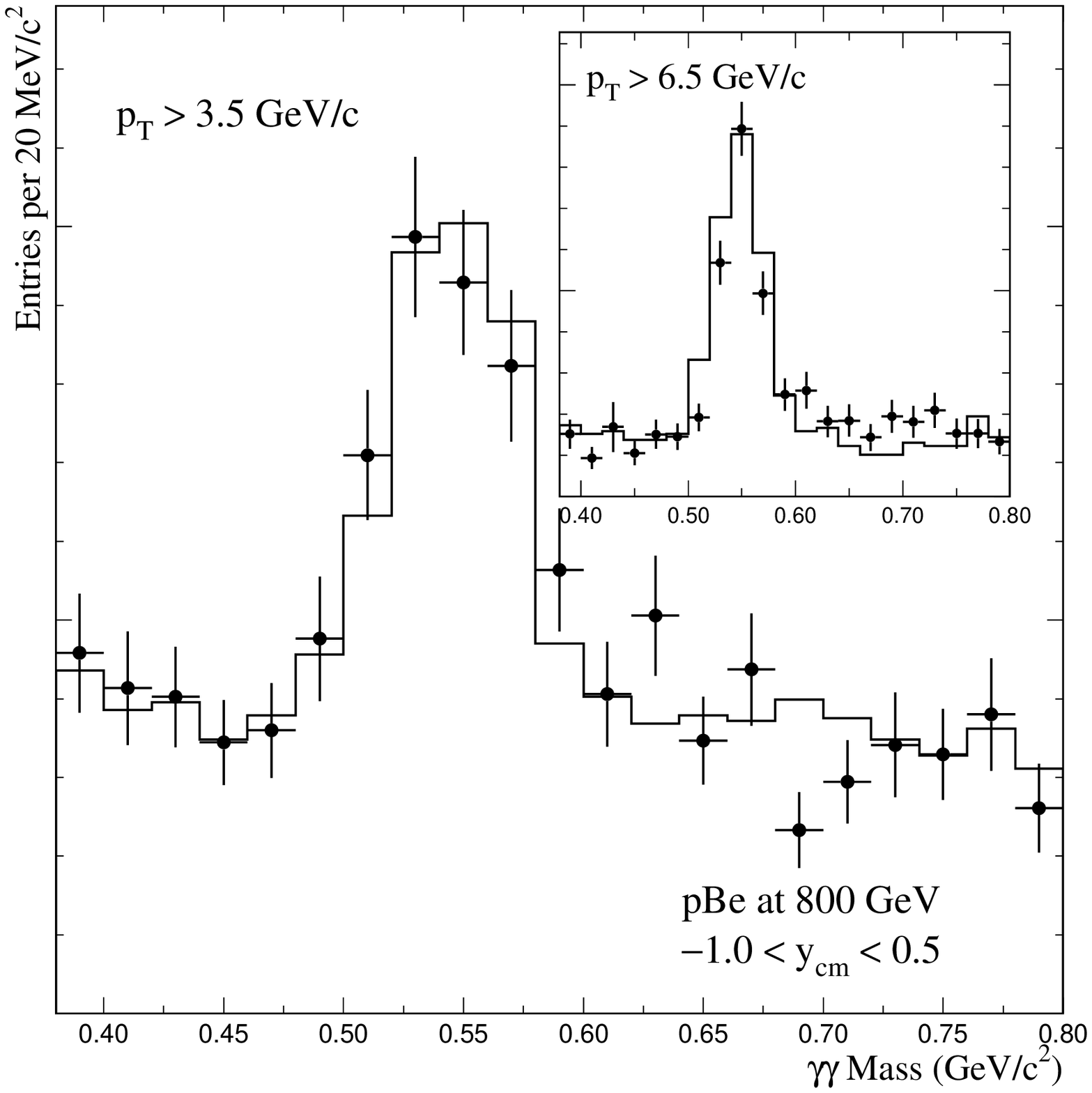}
\caption{
$\gamma\gamma$ mass distributions in the $\eta$ signal region in the 800 
GeV/$c$ data (histogram) compared to Monte Carlo simulations (points) for two 
requirements on minimum $p_T$.}
\label{fig:comp_mass_eta}
\end{figure}

\begin{figure}
\epsfxsize=5.5truein
\epsfbox[0 72 612 720]{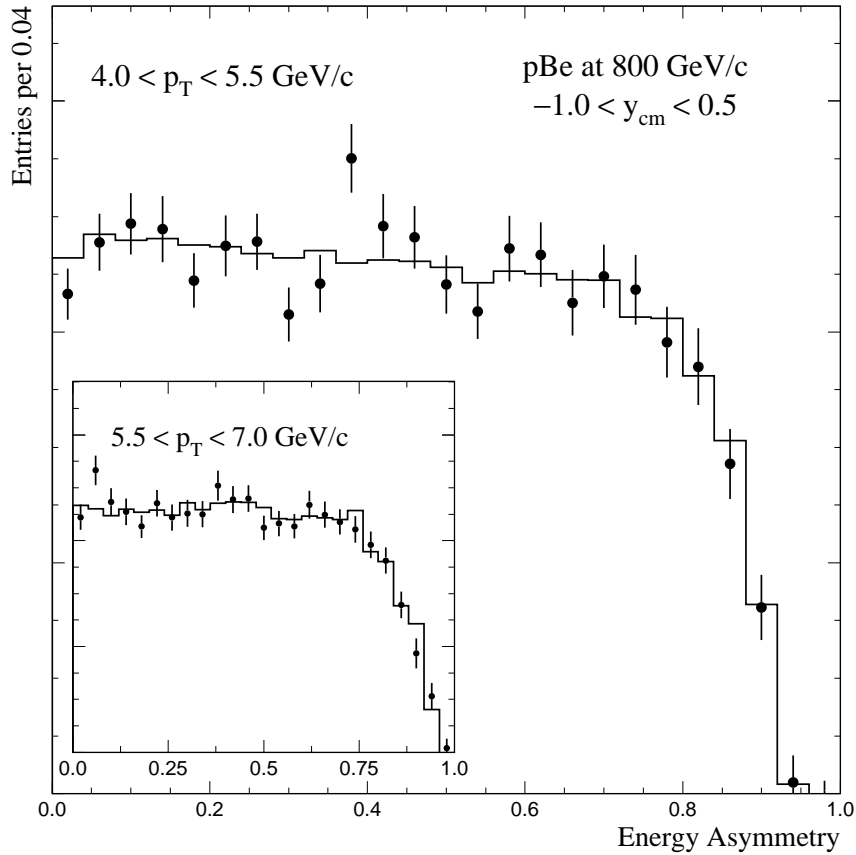}
\caption{
Comparison of $\gamma\gamma$ energy asymmetry distribution for \piz\ mesons 
in the data (histogram) and the Monte Carlo (points) for 800~GeV/$c$ \pBe\ 
interactions, for the $p_T$ intervals $4.0<p_T<5.5$~GeV/$c$ and 
$5.5<p_T<7.0$~GeV/$c$. These distributions have been corrected for 
contributions from background sources.}
\label{fig:asym}
\end{figure}

Comparisons between results from the Monte Carlo simulation and the
data for the distributions in $E_R-E_\Phi$ and in the fraction of
energy deposited in the front section of the EMLAC, \eft, are
presented in Figs.~\ref{fig:eref} and \ref{fig:efet}.  The Monte Carlo
results are in satisfactory agreement with the data, indicating that 
the simulation properly treats shower development in the EMLAC.
Figures~\ref{fig:comp_mass_pi0} and~\ref{fig:comp_mass_eta} show the
$\gamma\gamma$~mass spectra in the \piz\ and $\eta$ mass regions for
two minimum-$p_T$ requirements, and compare these spectra to the Monte
Carlo simulation. In addition to giving further evidence that the
Monte Carlo provides a good simulation of the resolution of the EMLAC,
the agreement in the levels of the combinatorial background indicates
that the Monte Carlo also provides reasonable simulation of the
underlying event structure.  Figure~\ref{fig:asym} shows a comparison
between the Monte Carlo and the data for the sideband-subtracted \piz\
energy asymmetry distribution, and the agreement indicates that the
Monte Carlo simulation describes accurately the losses of low-energy
photons.

\begin{figure}
\epsfxsize=5.5truein
\centerline{\epsffile{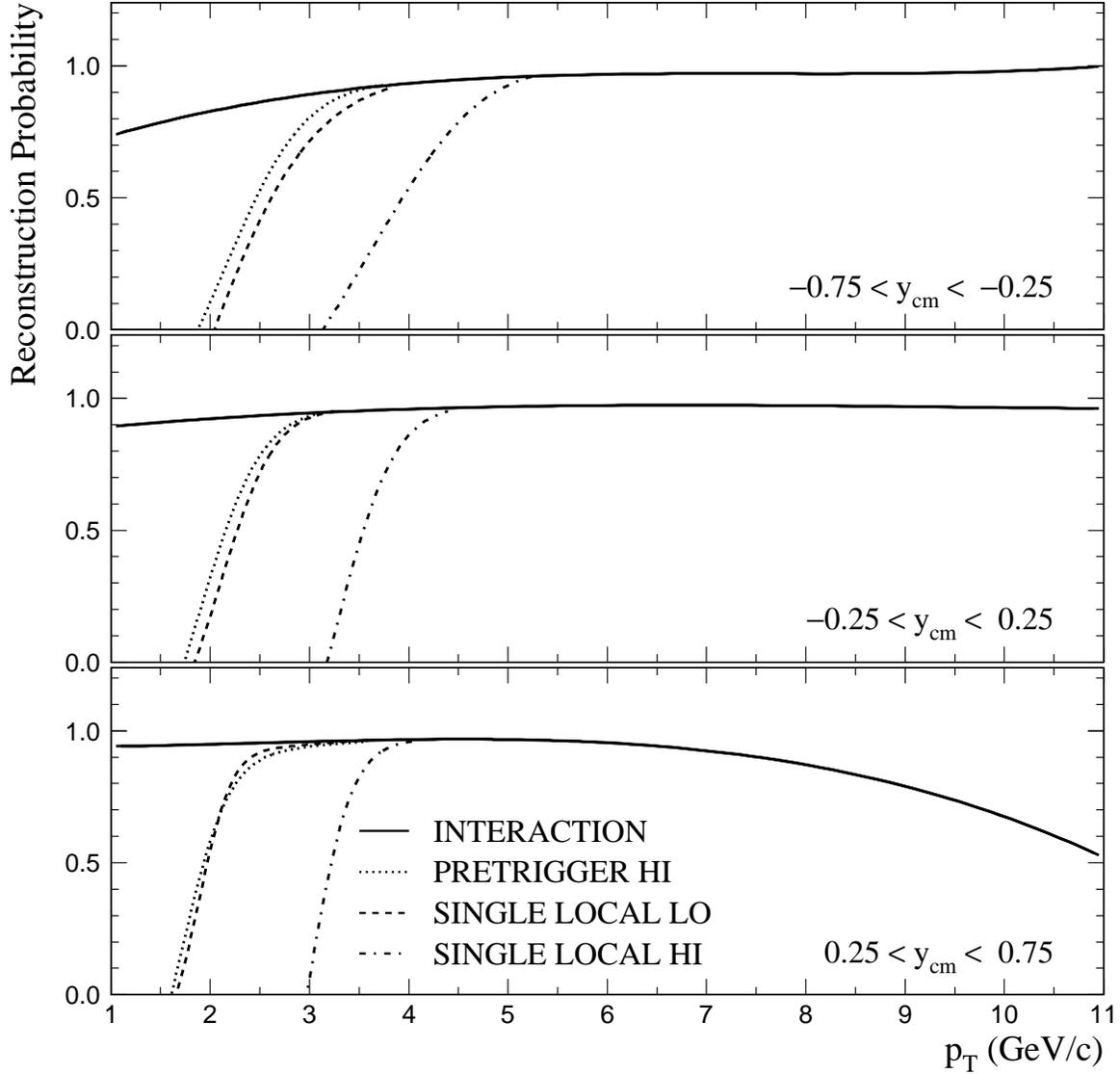}}
\caption{
Probability for accepting a \piz\ candidate (in the 530~GeV/$c$ Monte Carlo 
sample) as a function of $p_T$ for three rapidity intervals and different 
triggers. This probability reflects the losses due to the reconstruction 
algorithm, the \eft\ criterion, and the individual trigger requirements 
(see text for details).
}
\label{fig:receff}
\end{figure}

Reconstruction inefficiencies for \piz\ and $\eta$ mesons (which
satisfied the $A_{\gamma\gamma}$, energy asymmetry, requirement) were
relatively small over most of the kinematic range.
Figure~\ref{fig:receff} shows the probability for a \piz\ to pass the
selection requirements imposed on the Monte Carlo events at 530
GeV/$c$, as a function of $p_T$, for different rapidity intervals.
This probability includes losses due to the reconstruction algorithm,
the \eft\ requirement, and the 10$\%$ minimum trigger probability
requirement.  The inefficiency at forward rapidities and high-$p_T$ is
attributable to the increased difficulty in separating two photons
from \piz\ decays (coalescence) in this kinematic region.

\subsection{Normalization}

Electronic scalers that counted signals from the beam hodoscope,
interaction counters, and beam hole counters were used to determine
the number of beam particles incident on the target. Other scalers
logged the state of the trigger and of components of the data
acquisition system. Information from these scalers was used to
determine the number of beam particles that traversed the spectrometer
when it was ready to record data. This number was corrected for
multiple occupancy in the beam hodoscope (beyond that excluded via the
\BEAMONE\ requirement, and for absorption of beam in the target
material.

The normalization of the low \pt\ \piz\ cross section was
independently verified using events from the prescaled \BEAM\ and \INT\
trigger samples.  In these samples, the absolute normalization can be
obtained just by counting events. For these low \pt\ events, the
normalization as determined via the scalers and via event counting
techniques agree to better than~$3\%$.

Based upon the good agreement between results from these independent
normalization methods, combined with the stability of the cross
section results from various sections of the run, an evaluation of the
internal consistency of the scalers, and a detailed analysis of the
design, implementation and performance of the trigger, the net
systematic uncertainty in the overall normalization is $\approx$8$\%$.

\subsection{Secondary Beam Contamination}

The 530~GeV/$c$ secondary beam cross section measurements were
corrected for the small admixture of pions and kaons present in the
beam, estimated as 2.75$\%$ \pip\ and 0.5$\%$ \kpl,
respectively~\cite{striley}. Although the percentage contamination is 
small, it's effect at high \pt\ is enhanced by the presence of two, rather
than three, valence quarks in incident mesons.  The effect of \pip\
contamination was estimated using our high statistics study of \piz\
and $\eta$ meson production by 515~GeV/$c$ \pim\
beam~\cite{prl,piminus-prd}. This is justified because: neutral
meson production by \pip\ and \pim\ beams is expected to be similar,
based on isospin arguments and earlier measurements~\cite{bonesini};
the \pip\ component of the nominally 530~GeV/$c$ positive secondary
beam had a mean momentum of 515~GeV/$c$; and the ratio of the measured
high \pt\ \piz\ cross section from the \v{C}erenkov-tagged \pip\
component of the positive secondary beam to the corresponding cross
section from the 515~GeV/$c$ \pim\ beam was consistent with
unity~\cite{striley}.

The effect of \kpl\ contamination was assumed to be half that of the
same amount of \pip\ contamination, consistent with earlier, lower
energy measurements~\cite{donaldson_pi0} and with our own, more
statistically limited, data~\cite{striley}.  After correcting for beam
contamination, the 530~GeV/$c$ cross sections were reduced by
$\approx 2\%$ at low \pt\ and by $\approx 10\%$ at high \pt.

\subsection{Summary of systematic uncertainties}

The principal contributions to the systematic uncertainty arose from
the following sources: calibration of photon energy response
$(5$--$9\%)$, \piz\ and $\eta$ reconstruction efficiency and
detector-resolution unsmearing $(5\%)$, the overall normalization
$(8\%)$, and, for the 530~GeV/$c$ secondary beam, the beam
contamination $(0$--$7\%)$.  A more complete list of systematic
uncertainties is presented in Table~\ref{syserrs}.  Some of these
uncertainties ($e.g.$ normalization) are strongly correlated between
bins. The systematic uncertainties, combined in quadrature, are quoted
with cross sections in the appropriate tables.  Combining all
systematic uncertainties yields a net uncertainty that varies from
$11\%$ ($13\%$) at low \pt\ for \piz\ ($\eta$) mesons to $15\%$
at high-\pt.

The secondary proton beam was determined to have a mean momentum of
$530~\pm~2~{\rm GeV}/c$ with an estimated halfwidth of $\approx
30~{\rm GeV}/c$. This momentum spread introduces a small uncertainty
($\approx 5\%$) in comparisons of theory with data. (For the 800~GeV
primary beam, the momentum bite and the corresponding uncertainty are
negligible.)

\section{INCLUSIVE CROSS SECTIONS}

\subsection{\piz\ production}

\begin{figure}
\epsfxsize=5.5truein
\centerline{\epsffile{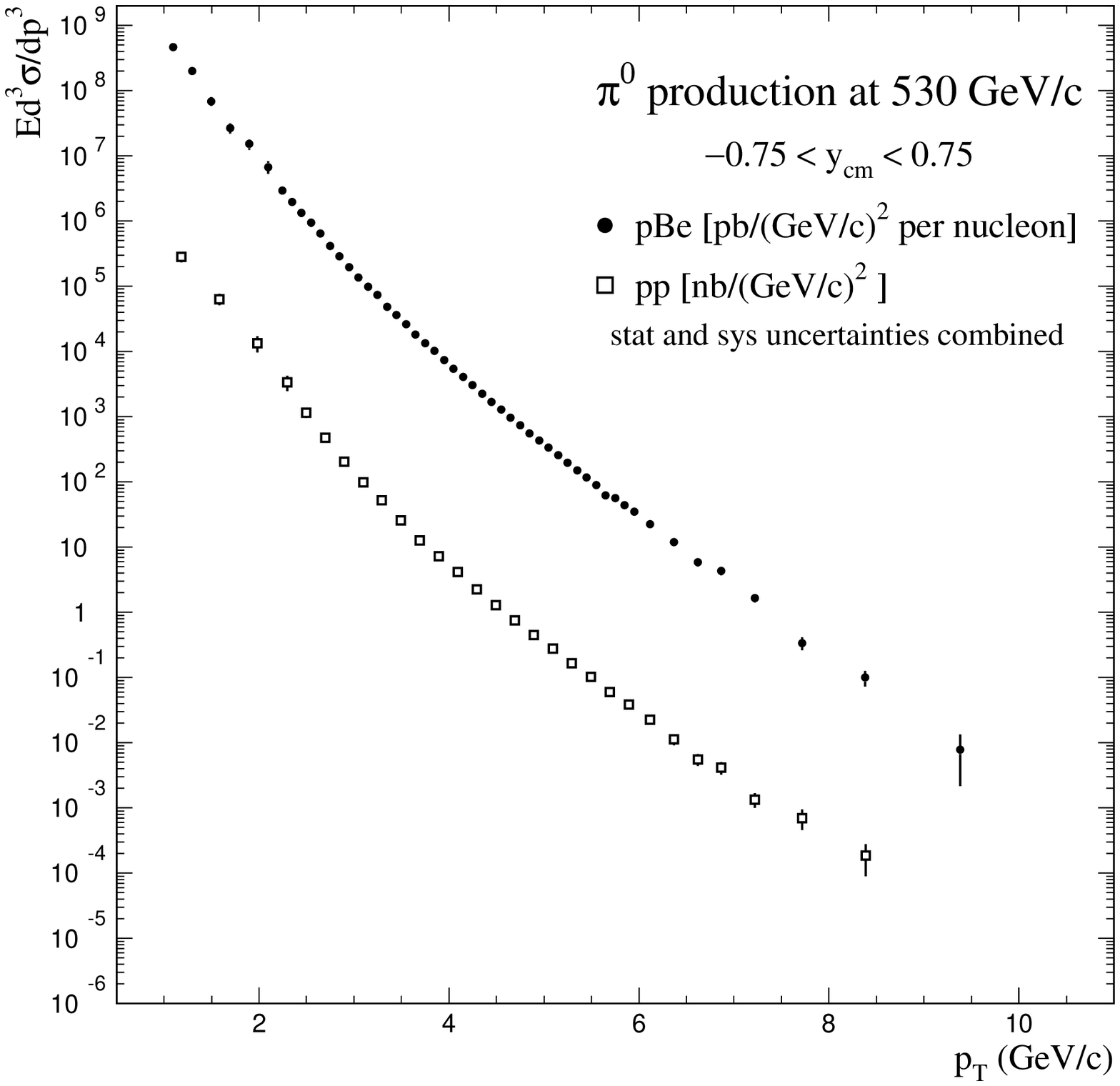}}
\caption{
Invariant differential cross sections (per nucleon) for \piz\ production 
as a function of \piz\ \pt\ in \pp\ and \pBe\ interactions at 530~GeV/$c$.
Cross sections have been averaged over the full rapidity range, 
$-0.75 \, \le \, \ycm\ \, \le \, 0.75$. The error bars have experimental 
statistical and systematic uncertainties added in quadrature.
\label{fig:pixs_pt_530}}
\end{figure}

\begin{figure}
\epsfxsize=5.5truein
\centerline{\epsffile{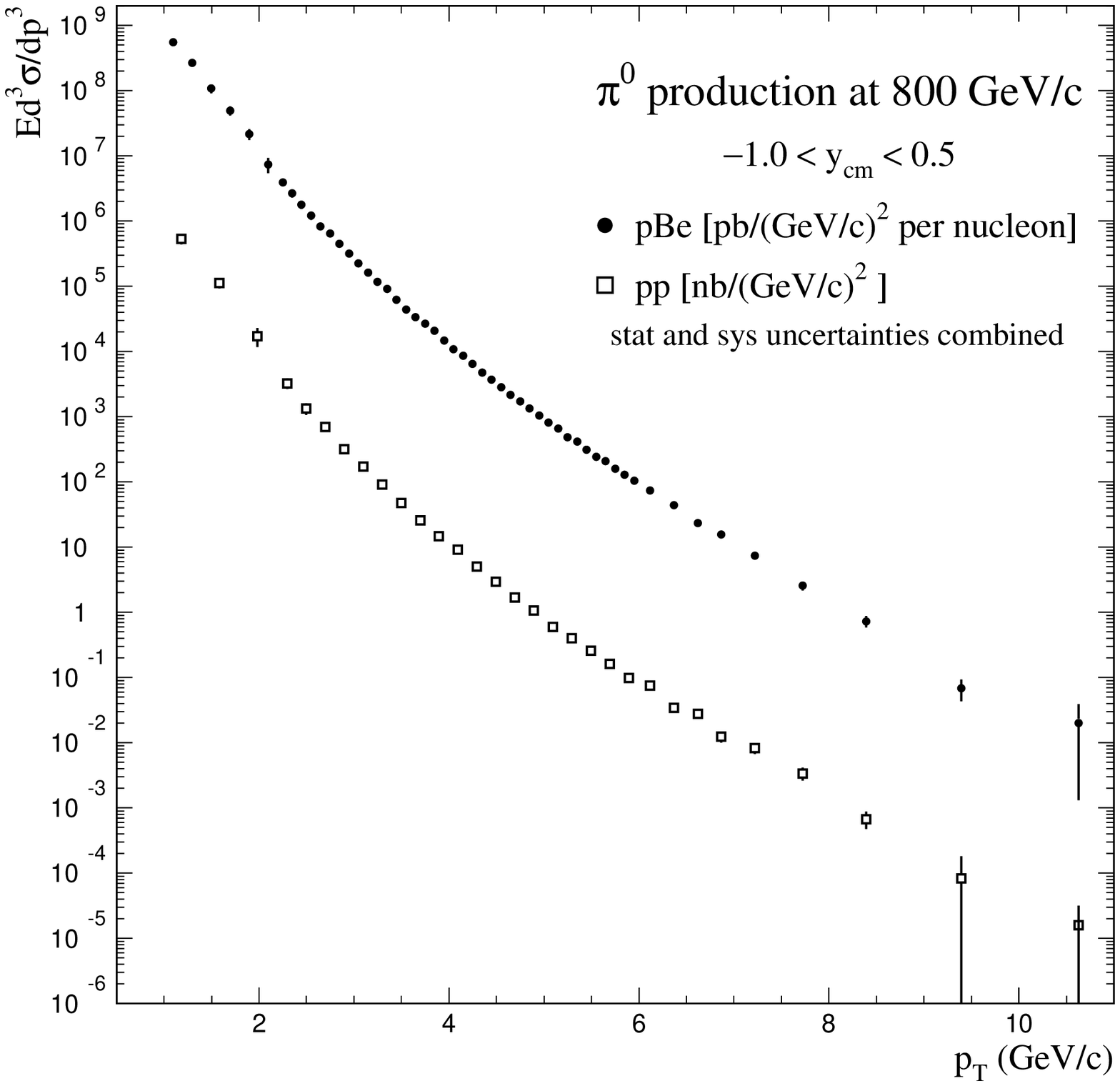}}
\caption{
Invariant differential cross sections (per nucleon) for \piz\ production 
as a function of \piz\ \pt\ in \pp\ and \pBe\ interactions at 800~GeV/$c$.
Cross sections have been averaged over the full rapidity range, 
$-1.0 \, \le \, \ycm\ \, \le \, 0.5$. The error bars have experimental 
statistical and systematic uncertainties added in quadrature.
\label{fig:pixs_pt_800}}
\end{figure}

The inclusive \piz\ cross sections per nucleon versus \pt\ are shown
in Figs.~\ref{fig:pixs_pt_530} and \ref{fig:pixs_pt_800} for 530 and
800~GeV/$c$ proton beams, respectively, incident upon beryllium and
liquid hydrogen targets. Because of the steeply falling spectra,
the data are plotted at abscissa values that correspond to the average 
values of the cross section in each \pt\ bin (assuming local exponential 
\pt\ dependence) \cite{laff}. These cross sections are also tabulated in 
Tables~\ref{pi0_table_be} and \ref{pi0_table_p}. The corresponding
cross section measurements, as functions of \pt\ and \ycm\ are reported
in Tables~\ref{pi0_table_be_530_rap} through \ref{pi0_table_p_800_rap}.

\begin{figure}
\epsfxsize=5.5truein
\centerline{\epsffile{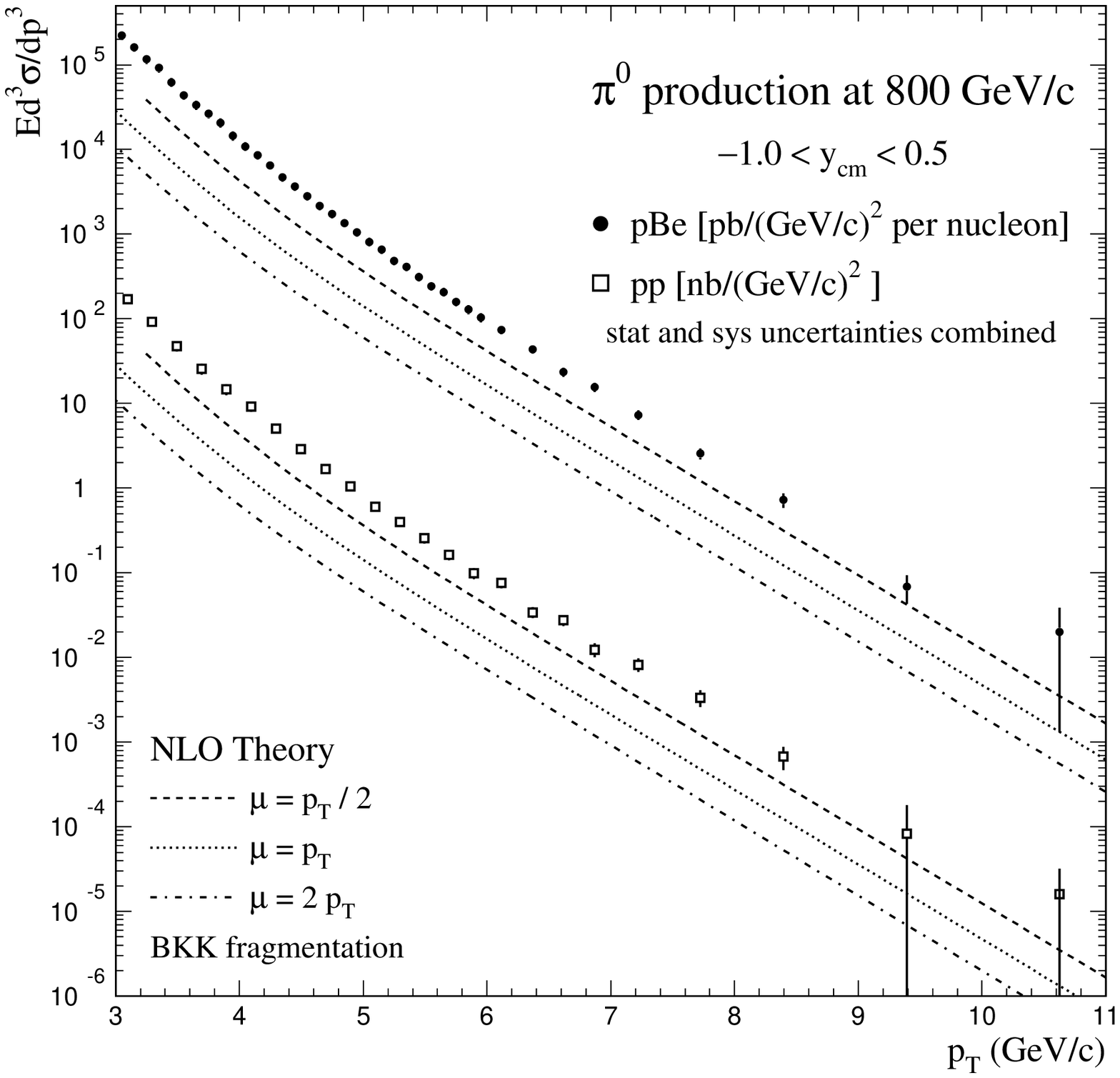}}
\caption{
Invariant differential cross sections (per nucleon) for \piz\ production 
as a function of \piz\ \pt\ in \pBe\ and \pp\ interactions at 800~GeV/$c$ 
compared to NLO PQCD calculations, with scale choices of 
$\mu$ = $\pt/2$, $\pt$, and $2\pt$. The error bars have experimental 
statistical and systematic uncertainties added in quadrature.
\label{fig:pixs_pt_800_qsq}}
\end{figure}

\begin{figure}
\epsfxsize=5.5truein
\centerline{\epsffile{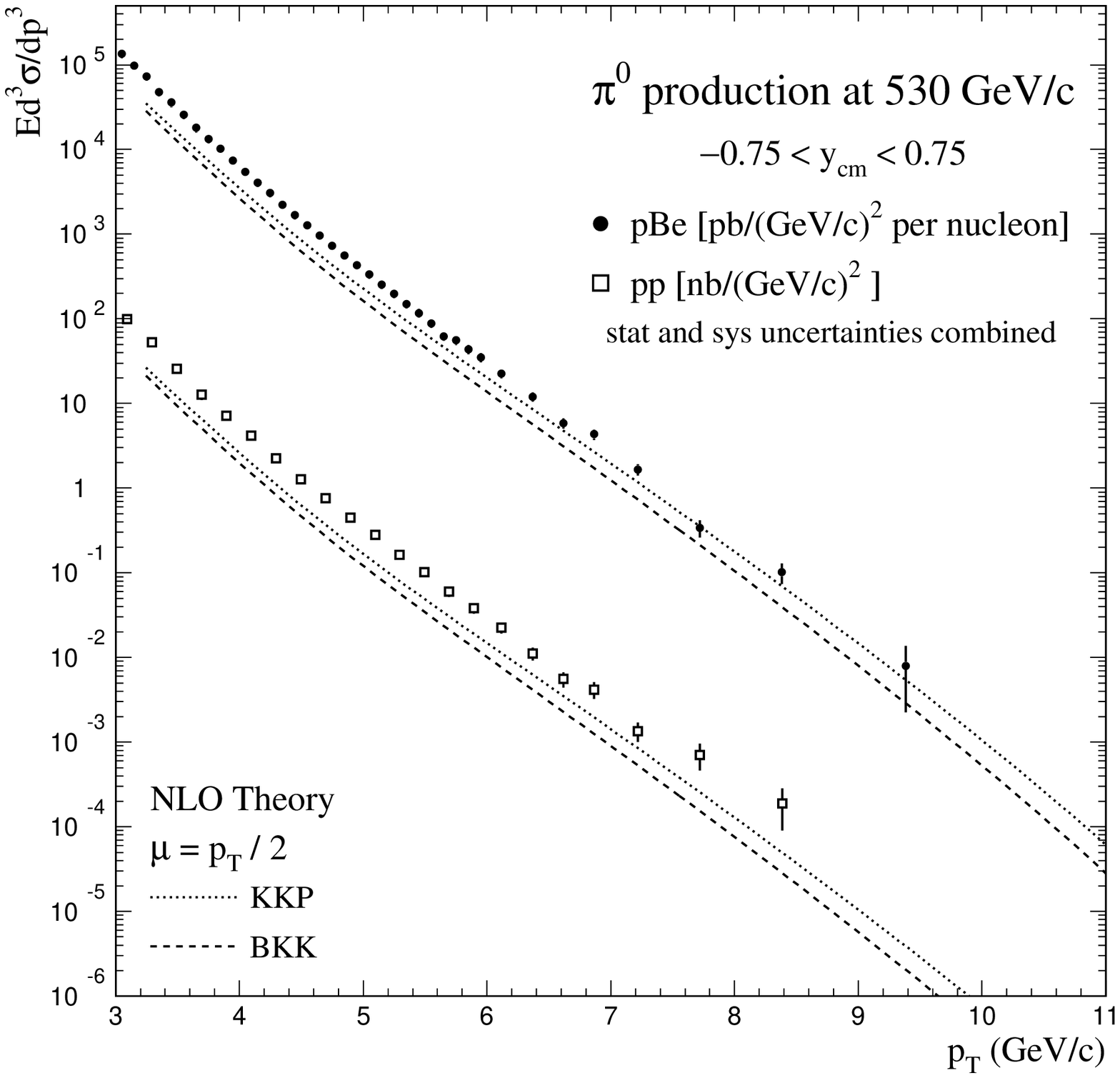}}
\caption{
Invariant differential cross sections (per nucleon) for \piz\ production 
as a function of \piz\ \pt\ in \pBe\ and \pp\ interactions at 530~GeV/$c$ 
compared to NLO PQCD calculations, with scale $\mu$ = $\pt/2$ and BKK and 
KKP fragmentation functions. The error bars have experimental statistical
and systematic uncertainties added in quadrature.
\label{fig:pixs_pt_530_frag}}
\end{figure}

\begin{figure}
\epsfxsize=5.5truein
\centerline{\epsffile{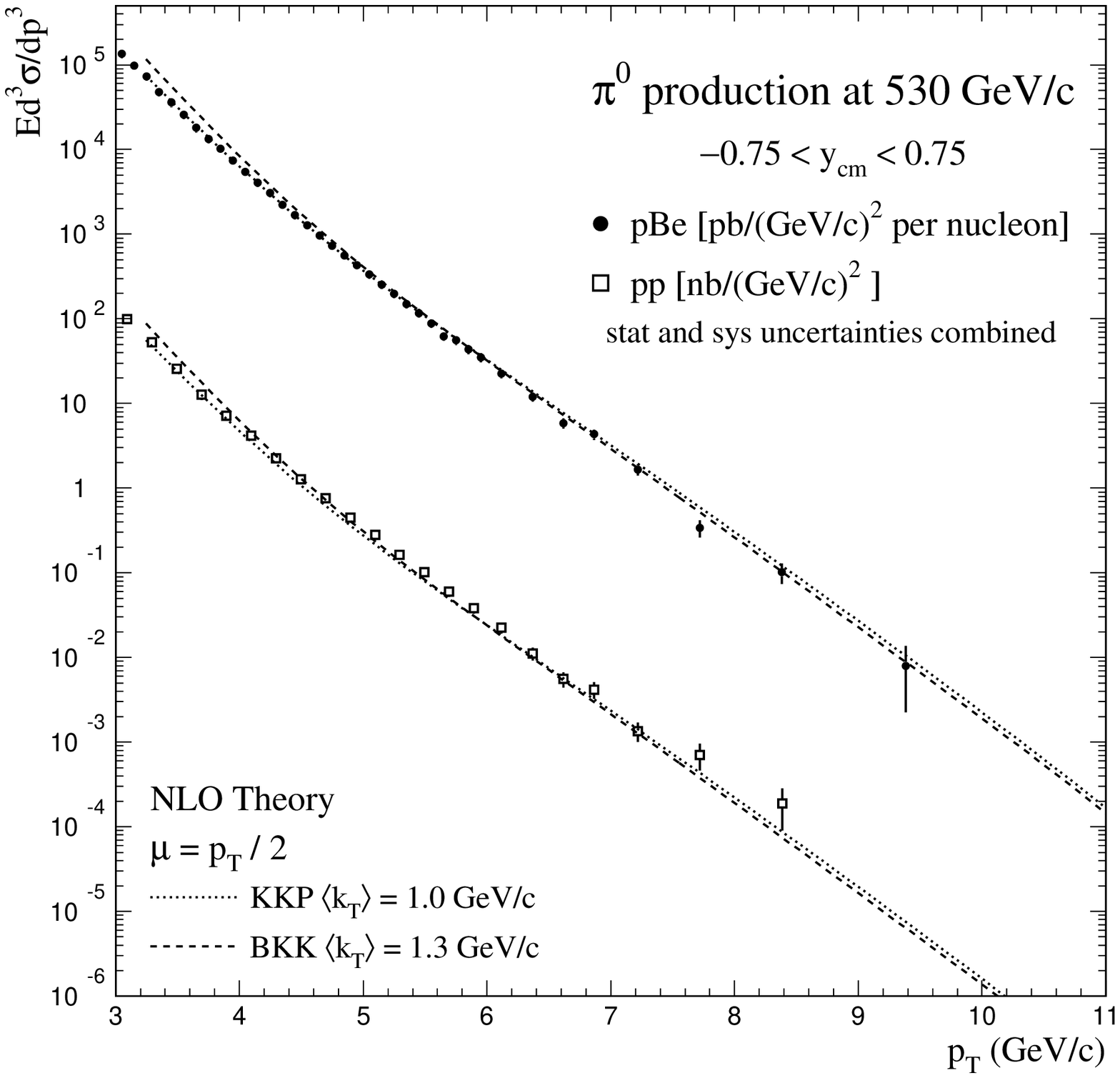}}
\caption{
Invariant differential cross sections (per nucleon) for \piz\ production 
as a function of \piz\ \pt\ in \pBe\ and \pp\ interactions at 530~GeV/$c$ 
compared to \kt-enhanced NLO PQCD calculations with scale $\mu$ = $\pt/2$.
Comparisons are shown for both BKK and KKP fragmentation functions.
The error bars have experimental statistical and systematic uncertainties
added in quadrature.
\label{fig:pixs_pt_530_kt}}
\end{figure}

\begin{figure}
\epsfxsize=5.5truein
\centerline{\epsffile{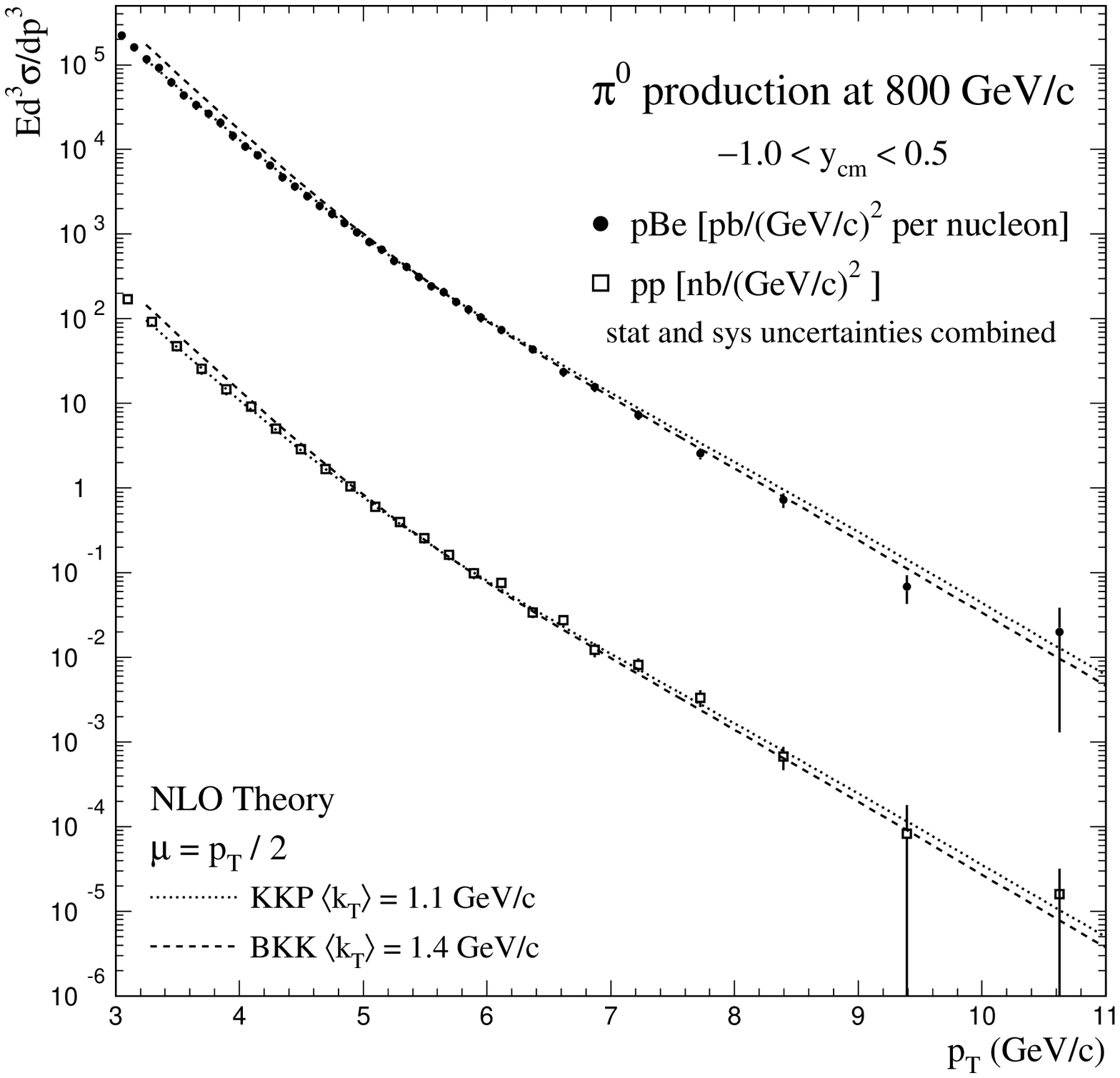}}
\caption{
Invariant differential cross sections (per nucleon) for \piz\ production 
as a function of \piz\ \pt\ in \pBe\ and \pp\ interactions at 800~GeV/$c$ 
compared to \kt-enhanced NLO PQCD calculations, with scale $\mu$ = $\pt/2$.
Comparisons are shown for both BKK and KKP fragmentation functions.
The error bars have experimental statistical and systematic uncertainties
added in quadrature.
\label{fig:pixs_pt_800_kt}}
\end{figure}

\begin{figure}
\epsfxsize=5.5truein
\centerline{\epsffile{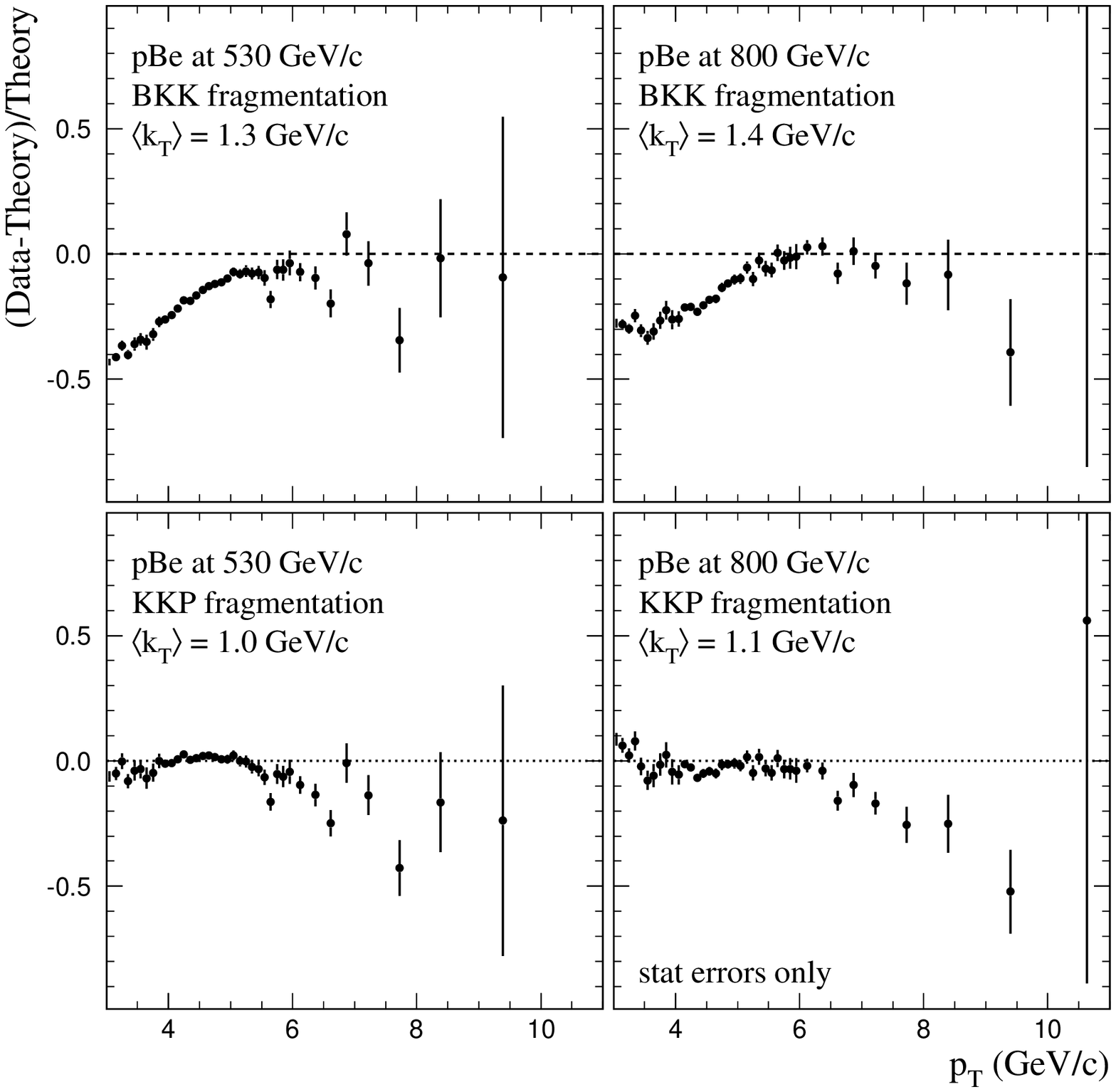}}
\caption{
Fractional difference between data and \kt-enhanced NLO PQCD calculations
for \piz\ production in \pBe\ interactions at 530 and 800~GeV/$c$ as a 
function of \piz\ \pt. The error bars represent only statistical 
contributions.
\label{fig:pixs_be_2x2}}
\end{figure}

\begin{figure}
\epsfxsize=5.5truein
\centerline{\epsffile{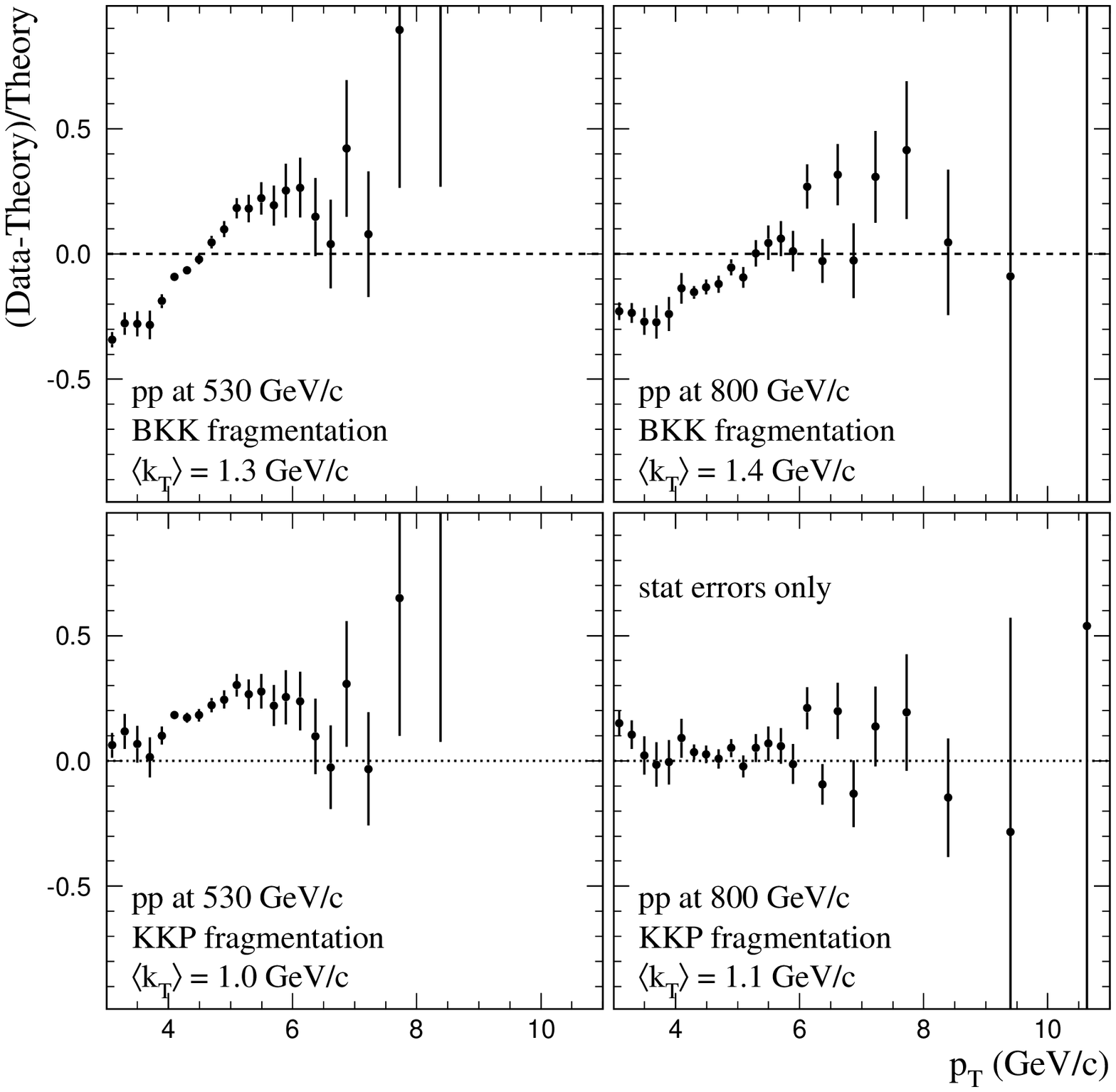}}
\caption{
Fractional difference between data and \kt-enhanced NLO PQCD results for
\piz\ production in \pp\ interactions at 530 and 800~GeV/$c$ as a function
of \piz\ \pt. The error bars represent only statistical contributions.
\label{fig:pixs_p_2x2}}
\end{figure}

\begin{figure}
\epsfxsize=5.5truein
\centerline{\epsffile{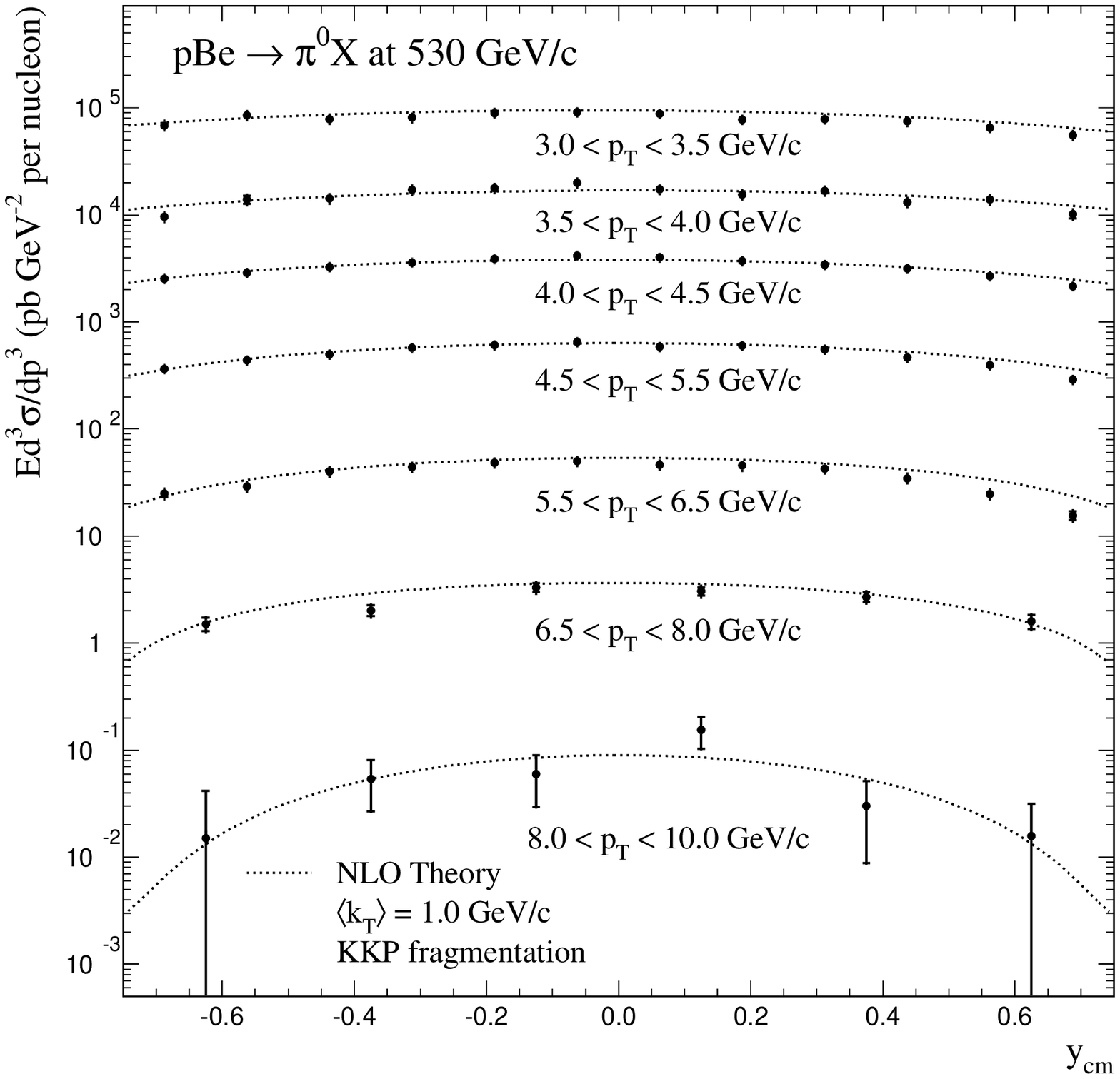}}
\caption{
Invariant cross sections per nucleon for \piz\ production in \pBe\ interactions
at 530~GeV/$c$. Cross sections are shown versus \ycm\ for several intervals 
in \pt. The curves represent the \kt-enhanced NLO QCD calculations for 
\avkt=1.0~GeV/$c$ and $\mu =\pt/2$, using KKP fragmentation functions.
The error bars have experimental statistical and systematic uncertainties
added in quadrature.
\label{fig:pixs_rap_530}}
\end{figure}

\begin{figure}
\epsfxsize=5.5truein
\centerline{\epsffile{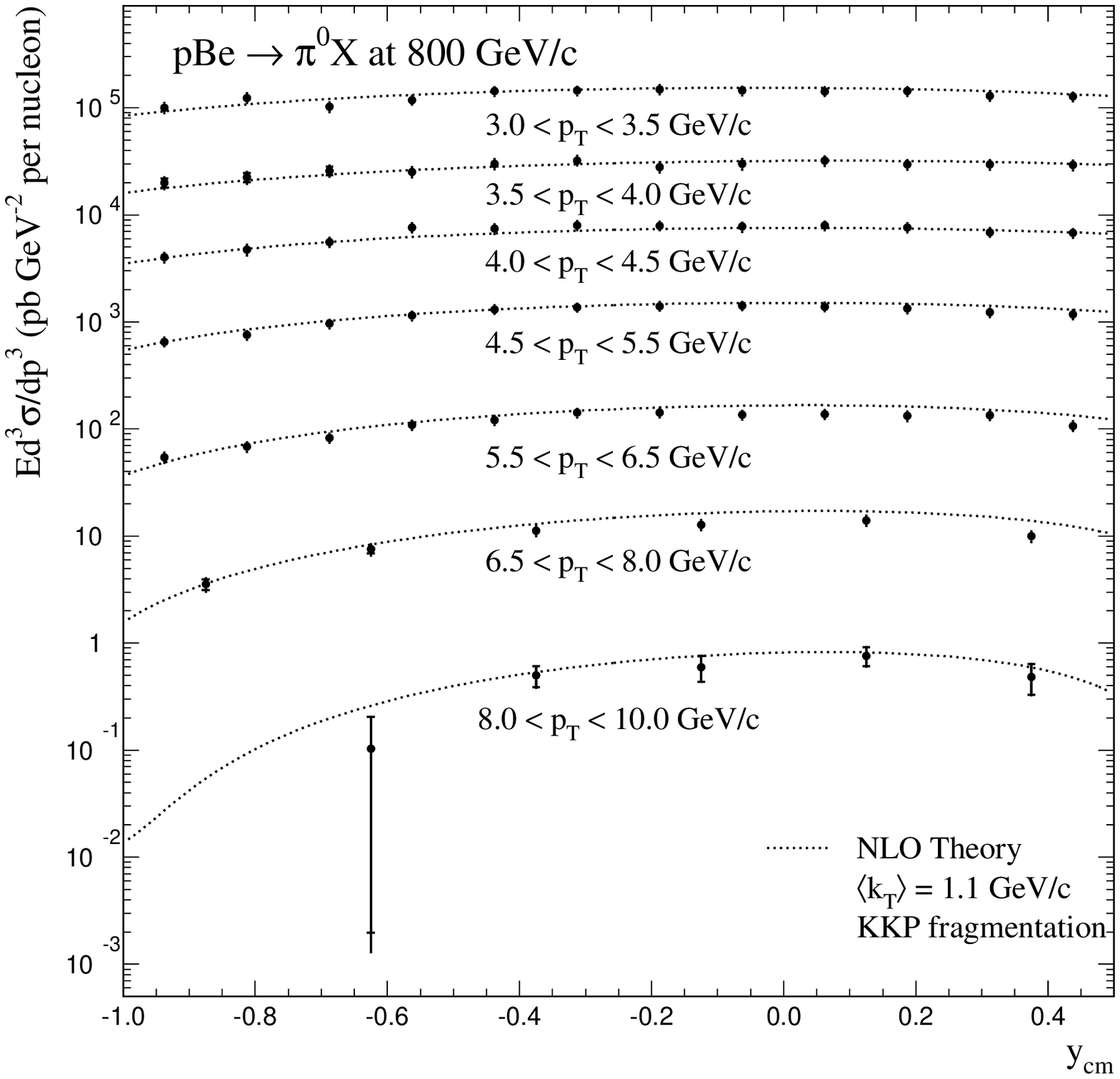}}
\caption{
Invariant cross section per nucleon for \piz\ production in \pBe\ interactions
at 800~GeV/$c$. Cross sections are shown versus \ycm\ for several intervals 
in \pt. The curves represent the \kt-enhanced NLO QCD calculations for
\avkt=1.1~GeV/$c$ and $\mu = \pt/2$, using KKP fragmentation functions.
The error bars have experimental statistical and systematic uncertainties
added in quadrature.
\label{fig:pixs_rap_800}}
\end{figure}

\begin{figure}
\epsfxsize=5.5truein
\centerline{\epsffile{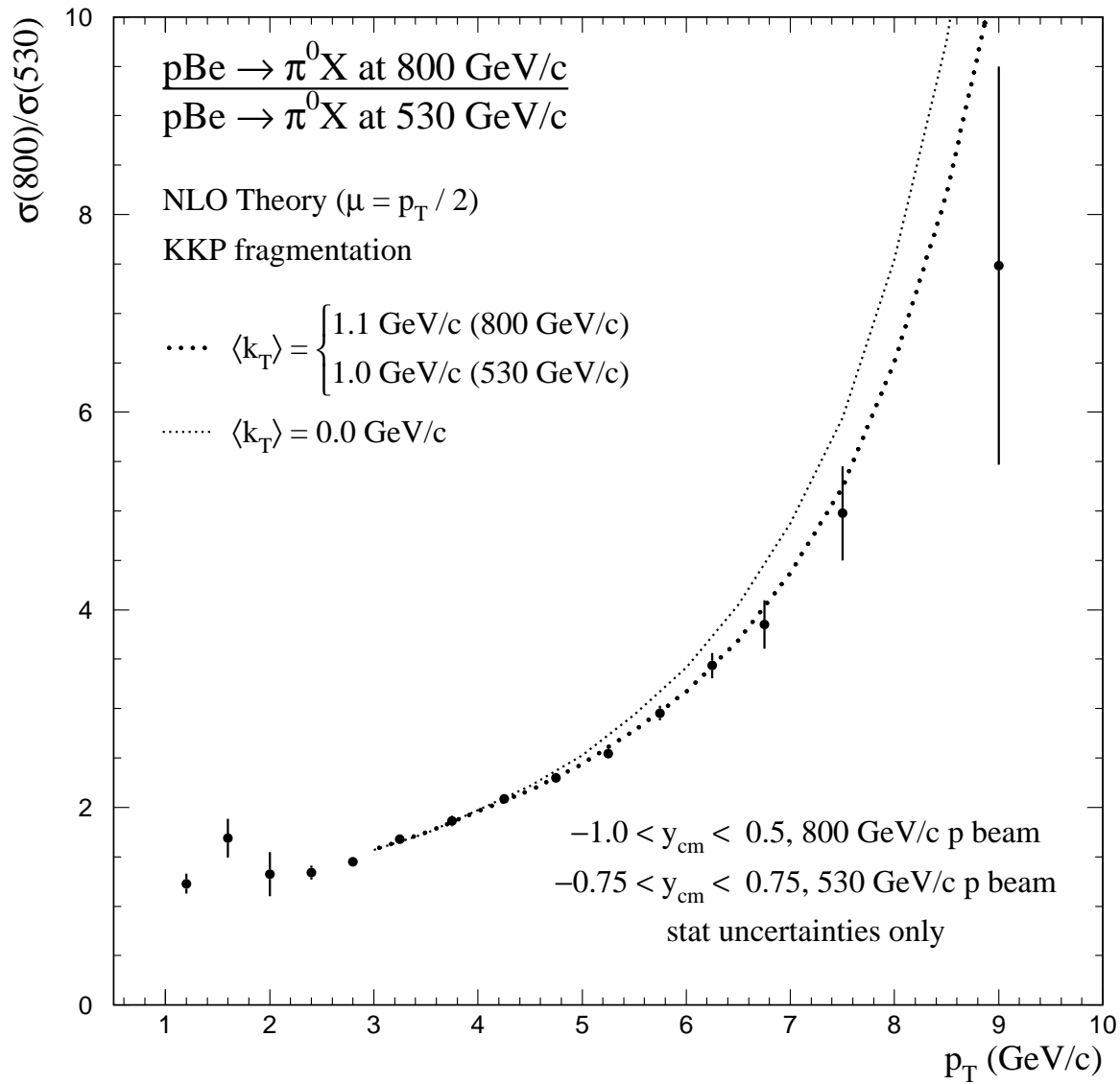}}
\caption{
Ratio of invariant cross sections for \piz\ production in \pBe\ interactions
at 800 and 530~GeV/$c$ as a function of \piz\ \pt, compared to conventional 
and \kt-enhanced NLO QCD results using KKP fragmentation functions. 
The error bars reflect only the statistical uncertainties.
\label{fig:pixs_pt_edep_be}}
\end{figure}

NLO PQCD calculations \cite{aversa} are compared to the data in
Figs.~\ref{fig:pixs_pt_800_qsq} through \ref{fig:pixs_pt_edep_be}. In
Fig.~\ref{fig:pixs_pt_800_qsq}, NLO PQCD results using CTEQ4M parton
distribution functions \cite{cteq4} and BKK fragmentation functions
\cite{bkk} are compared to the measured inclusive \piz\ cross sections
for \pBe\ and \pp\ interactions at 800~GeV/$c$.  The PQCD calculations
for the Be target have been adjusted to account for nuclear effects by
the factor $A^{\alpha-1}$, where $A$ is the atomic number and $\alpha
= 1.12~(1.08)$ for 530 (800)~GeV/c incident protons. Theoretical
results are presented for three values of the factorization scale:
$\mu$ = $\pt/2$, $\pt$, and $2\pt$. (In these comparisons, the
renormalization and fragmentation scales have been set to the value of
the factorization scale.)  In addition to a large dependence on the
choice of scale, the expectations for these choices of $\mu$ lie
significantly below the data, at both 530 and 800~GeV/$c$.

In Fig.~\ref{fig:pixs_pt_530_frag}, NLO calculations using BKK and KKP
\cite{kkp} fragmentation functions, and $\mu$ = $\pt/2$, are compared
to the \piz\ cross sections for 530~GeV/$c$ \pBe\ and \pp\
interactions.  The calculations exhibit considerable dependence on the
choice of fragmentation function, but both choices predict yields that
are significantly lower than the data.

These discrepancies have been interpreted \cite{prl,prd,Huston} as 
arising from additional soft-gluon emission in the initial state that 
is not included in the NLO calculation, and which results in sizeable 
parton \kt\ before the hard collision (for a different perspective, see 
the discussion in Ref.~\cite{frenchpiz}). Soft-gluon (or \kt) effects 
are expected in all hard-scattering processes, such as the inclusive
production of jets, high-\pt\ mesons, and direct
photons~\cite{FF,font,cont1,cont2}.  The Collins-Soper-Sterman
resummation formalism \cite{css} provides a rigorous basis for
understanding these radiation effects, and there have been several
recent efforts to derive resummation descriptions for the inclusive
direct-photon \cite{Catani:1999hs,Catani,Lai:1998xq,sterman-res} and
dijet cross sections \cite{Laenen,Kidonakis,Kidonakis2}. The
calculation of Ref.~\cite{Catani:1999hs} for inclusive direct-photon
production, which includes the effects of soft-gluon resummation near
the kinematic threshold limit
$\left( x_T = 2 p_T/ \sqrt{s} \longrightarrow 1 \right)$,
has a far smaller sensitivity to scale, compared to NLO calculations,
and provides cross sections close to those of NLO calculations with a
scale of $\mu = p_T/2$. Also, for our energies, the calculation of
Ref.~\cite{fink} using \kt-resummation, and of
Ref.~\cite{sterman-res}, which treats simultaneously threshold and
recoil effects in direct-photon production, yield a substantially
larger cross section than the NLO result.  However, no such
calculations are available for inclusive meson production.  In their
absence, we use a PQCD-based model that incorporates transverse
kinematics of initial-state partons to study the principal
consequences of additional \kt\ for high-\pt\ production processes.

Because the unmodified PQCD cross sections fall rapidly with 
increasing \pt, the net effect of the ``\kt\ smearing'' is to increase 
the expected yield at higher \pt's. An exact treatment of the modified 
parton kinematics has been implemented in a Monte Carlo calculation of the 
leading-order (LO) cross sections for high-\pt\ particle 
production \cite{owensll}, with the
\kt\ distribution for each of the incoming partons represented by a 
Gaussian with one adjustable parameter (\avkt).  Unfortunately, no
such program is available for NLO calculations, and so we approximate
the effect of \kt\ smearing by multiplying the NLO cross sections by
the corresponding LO ${k}_{T}$-enhancement factors.  Admittedly, this
procedure involves a risk of double-counting since some of the
${k}_{T}$-enhancement may already be contained in the NLO
calculation. However, we expect such double-counting effects to be
small.

The \avkt\ values used in the calculation of the LO \kt-enhancement
factors are similar to those employed in comparisons of kinematic
distributions in data involving production of high-mass
$\gamma\gamma$, $\gamma$\piz, and \piz\piz\ systems with calculations
relying on the same LO program (see Refs.~\cite{prl,prd} for further
details). For these comparisons, we used the LO versions of the CTEQ4
distribution and (where appropriate) BKK fragmentation functions, and
an average transverse momentum of 0.6~GeV/$c$ for the \piz\ mesons
relative to the fragmenting parton direction (varying this parameter
in the range 0.3 to 0.7~GeV/$c$ does not affect our conclusions)
\cite{R702-jets,CCOR-qt}.

Comparisons of the \kt-enhanced calculations with data at 530~GeV/$c$
are displayed in Fig.~\ref{fig:pixs_pt_530_kt}, indicating reasonable
agreement for the \avkt\ values chosen. Similar conclusions can be
drawn from comparisons between calculations and data at 800~GeV/$c$,
as illustrated in Fig.~\ref{fig:pixs_pt_800_kt}.  It is interesting to
compare the fractional differences between data and the \kt-enhanced
NLO calculations using BKK and KKP fragmentation functions on a linear
scale.  Such comparisons are shown in Fig.~\ref{fig:pixs_be_2x2} for
\pBe\ interactions at 530 and 800~GeV/$c$.  The \kt-enhanced
calculations using the KKP fragmentation functions are seen to reproduce
the shape of the \piz\ cross section better than calculations
using the earlier BKK fragmentation functions, but small discrepancies
are still present.  Similar conclusions can be drawn from comparisons
of NLO QCD with our data on \pp\ collisions
(Fig.~\ref{fig:pixs_p_2x2}), although in this case the fractional
differences using the KKP fragmentation functions are systematically
greater than zero when we employ the same \avkt\ values used in
Fig.~\ref{fig:pixs_be_2x2}. It is, however, worth noting that the
hydrogen nucleus is not typical, having an up to down quark ratio 
of two rather than a value of less than one as in 
the case of standard nuclei, such as Be (Cu) for which $\mit{u/d}$ =
0.93 (0.94). The theoretical calculations address this difference by
calculating the nuclear cross sections using the actual numbers of
protons and neutrons present in each case, but this alone does not
fully compensate for the influence of the nuclear media surrounding
interactions in targets other than hydrogen.  As a result, values of
\avkt\ that yield agreement with nuclear data adjusted by $A^{\alpha-1}$ 
may not be fully appropriate for hydrogen.  In particular, raising
\avkt\ by only 0.05~GeV/$c$ drops the points in the lower two
quadrants of Fig.~\ref{fig:pixs_p_2x2} by enough ($\approx 0.075$ at
4.5~GeV/c) to approximately center these distributions on zero.

Figures~\ref{fig:pixs_rap_530} and \ref{fig:pixs_rap_800} show the
cross sections for inclusive \piz\ production as functions of rapidity
for \pBe\ interactions at 530 and 800~GeV/$c$, for several intervals
in \pt.  The expected peaking at a scattering angle near $90^\circ$ in
the center of mass (\ycm\ = 0) develops slowly as a function of \pt.
The shapes and normalizations of the data are in good agreement with
the \kt-enhanced calculations.

Both theoretical and experimental uncertainties are reduced in the ratio 
of invariant cross sections for \piz\ production at 800 and 530~GeV/$c$, 
allowing, in principle, a more sensitive test of the calculations. 
Figure~\ref{fig:pixs_pt_edep_be} displays this ratio compared to
conventional (\avkt=0) and \kt-enhanced NLO results using KKP 
fragmentation functions.  The \avkt\ values used here correspond
to those used in Figs.~\ref{fig:pixs_pt_530_kt} and
\ref{fig:pixs_pt_800_kt}.  The energy dependence of the data is
accommodated better by the \kt-enhanced theory. Similar conclusions
are obtained for the hydrogen target data (not shown).

The results discussed in this section are not very sensitive to the
reasonably well-known parton distribution functions \cite{prl} (quark
distributions being of primary importance here). Methods similar to 
the ones described in this paper have been applied to analyze 
high-\pt\ hadron spectra in \pp, \pA, and AA collisions \cite{wang2,zhang}, 
and \kt\ effects have been found important for describing data on 
inclusive production of charged mesons.

\subsection{$\eta$ production}

\begin{figure}
\epsfxsize=5.5truein
\centerline{\epsffile{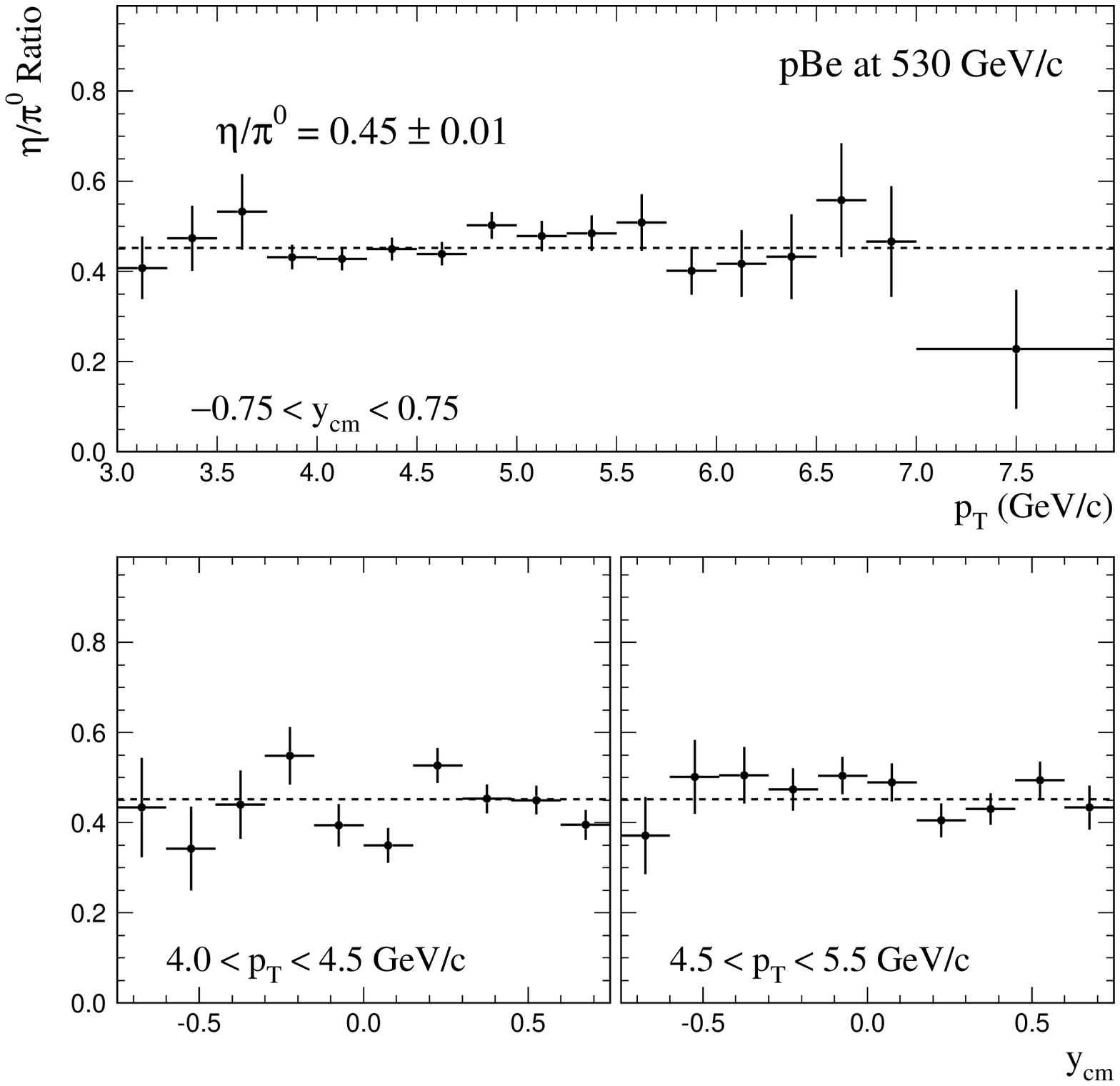}}
\caption{
Ratios of  $\eta$ to \piz\ invariant cross sections from 530~GeV/$c$ 
\pBe\ interactions, as functions of \pt\ (top), and of \ycm\ (bottom)
for two \pt\ ranges. The error bars reflect only the statistical uncertainties.
\label{fig:etapi_530}}
\end{figure}

\begin{figure}
\epsfxsize=5.5truein
\centerline{\epsffile{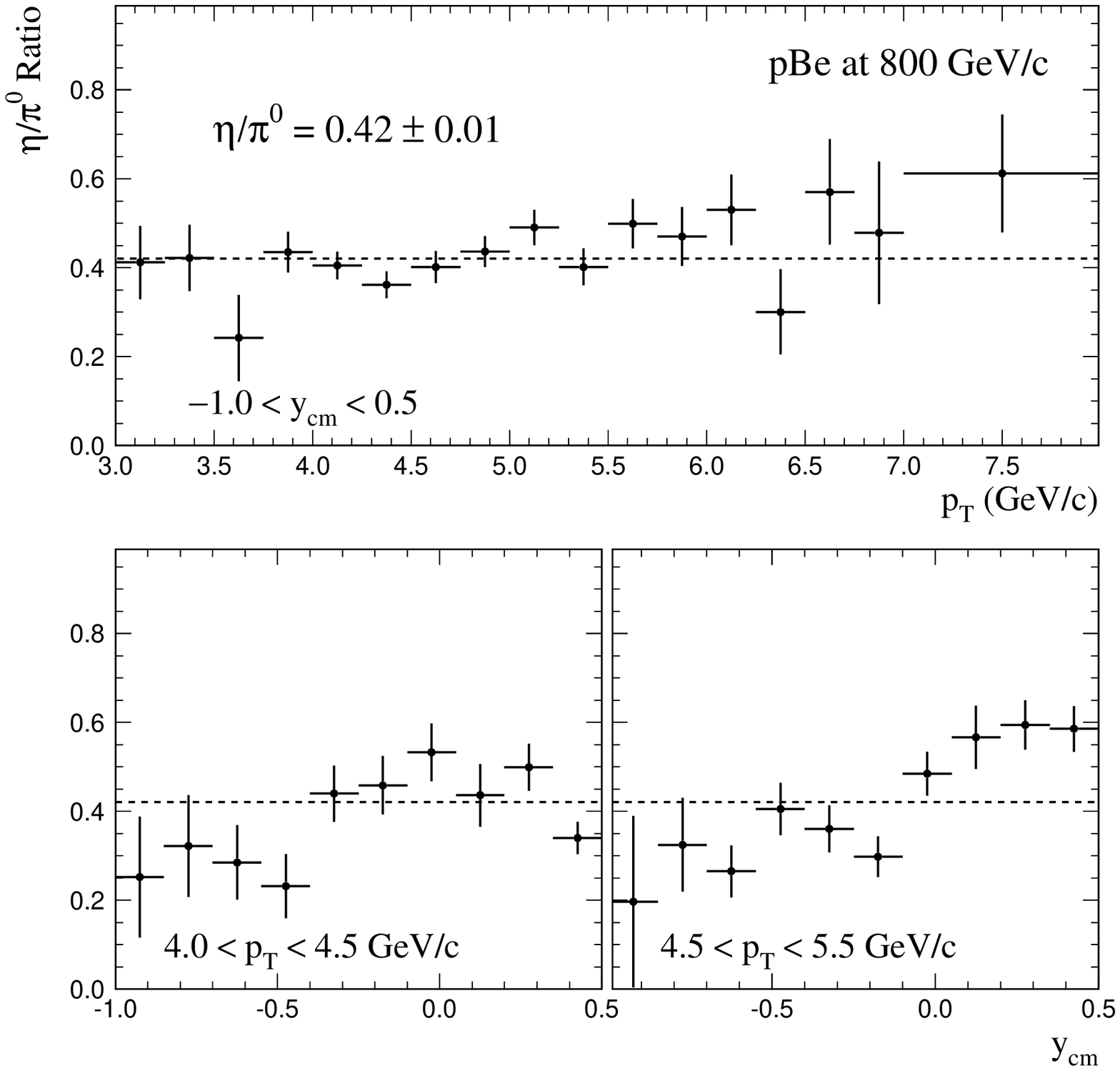}}
\caption{
Ratios of  $\eta$ to \piz\ invariant cross sections from 800~GeV/$c$
\pBe\ interactions, as functions of \pt\ (top), and of \ycm\ (bottom) for 
two \pt\ ranges. The error bars reflect only the statistical uncertainties.
\label{fig:etapi_800}}
\end{figure}

Cross sections for inclusive $\eta$ production are tabulated in
Tables~\ref{eta_table_be} through \ref{eta_table_p_800_rap}.
Theoretical descriptions of $\eta$-meson production differ from the
\piz\ case primarily because of differences in the fragmentation of
partons into the particles of interest. To investigate this aspect,
we present $\eta/\pi^0$ relative production rates as functions of
\pt\ and \ycm\ (for two \pt\ ranges) in Figs.~\ref{fig:etapi_530} and 
\ref{fig:etapi_800} --- the data average to a value of $0.45 \pm 0.01$ 
at 530~GeV/$c$, and $0.42 \pm 0.01$ at 800~GeV/$c$ 
(for 3 $< \, \pt\ \, < $ 8~GeV/$c$).

\section{SUMMARY}

The invariant cross sections for \piz\ and $\eta$ production have been
measured for \pp\ and \pBe\ collisions at 530 and 800~GeV/$c$ as
functions of \pt\ and \ycm, over the kinematic range $1~< \pt <~10$
GeV/$c$, and 1.5 units in rapidity. Results from \kt-enhanced NLO QCD
calculations are in reasonable agreement with our measured \piz\ cross
sections. Employing the recent KKP fragmentation functions in the NLO
QCD calculations was found to improve the description of the detailed
shape of our \piz\ cross sections (as a function of \pt) relative to
theoretical results obtained using the earlier BKK fragmentation
functions.  The measured $\eta/\pi^0$ production ratios, which provide
information about the relative fragmentation of partons into these
mesons, are $0.45 \pm 0.01$ at 530~GeV/$c$ and $0.42 \pm 0.01$ at 800
GeV/$c$, averaged over \pt\ and rapidity.

\acknowledgments

We thank the U.~S. Department of Energy, the National Science
Foundation, including its Office of International Programs and the
Universities Grants Commission of India, for their support of this
research.  The staff and management of Fermilab are thanked for their
efforts in making available the beam and computing facilities that
made this work possible.  We are also pleased to acknowledge the
contributions of our colleagues on Fermilab experiment E672. We
acknowledge the contributions of the following colleagues during the
upgrade and operation of the Meson West spectrometer:
W.~Dickerson, E.~Pothier from Northeastern University;
J.~T.~Anderson,
E.~Barsotti,~Jr.,
H.~Koecher,
P.~Madsen,
D.~Petravick,
R.~Tokarek, J.~Tweed, D.~Allspach, J.~Urbin, R.~L.~Schmitt and the cryo 
crews from Fermi National Accelerator Laboratory;
T.~Haelen, C.~Benson, 
L.~Kuntz, and D.~Ruggiero
from the University of Rochester;
the technical staffs of 
Michigan State University and
Pennsylvania State University for the construction of the straw tubes
and of the University of Pittsburgh for the silicon detectors. We
would also like to thank the following commissioning run collaborators 
for their invaluable contributions to the hardware and software
infrastructure of the original Meson West spectrometer: 
G.~Alverson,
G.~Ballocchi,
R.~Benson,
D.~Berg,
D.~Brown,
D.~Carey,
T.~Chand,
C.~Chandlee,
S.~Easo,
W.~Faissler,
G.~Glass,
I.~Kourbanis,
A.~Lanaro,
C.~A.~Nelson,~Jr.,
D.~Orris,
B.~M.~Rajaram,
K.~Ruddick,
A.~Sinanidis, and
G.~Wu.
We thank S.~Catani, J.Ph.~Guillet, B.~Kniehl, J.~Owens, 
G.~Sterman, and W.~Vogelsang for many helpful
discussions and for providing us with their QCD calculations.

%
%
\renewcommand{\baselinestretch}{1.}
\newpage
\bibliography{etapi}  

\begin{thebibliography}{10}
\bibitem{ssdnote}
These hybrid planes were installed after the commissioning run to provide
  improved vertex resolution.
\bibitem{htargnote}
The fiducial volume of the hydrogen target was defined to exclude interactions
  with the mylar walls of the target flask.
\bibitem{accnote}
Binned in \pt, rapidity, and vertex position, simulated \piz\ and $\eta$ mesons
  were allowed to decay isotropically to two photons. Photon pairs that
  satisfied the energy asymmetry requirement (as defined in the text) were
  projected to the EMLAC, and selected on the basis of the standard fiducial
  acceptance criteria.
\bibitem{massnote}
The calculations of invariant mass and \pt\ assumed that the photons originated
  from the reconstructed interaction vertex.
\bibitem{trignote}
The response of each trigger group was evaluated as a function of \pt\
  reconstructed within the group, using data samples that were not biased in
  that region. The sample used for obtaining the efficiency of the \SLH\
  trigger was based on information from octants that fired the \SLL\ trigger.
  The sample used to evaluate the response of the \SLL\ trigger used
  information from the octants opposite those that satisfied the \SLH\ trigger.
  In the case of the \PRETH, octants located opposite those that satisfied the
  \SLH\ trigger were used to measure the trigger efficiency for the trigger
  groups comprised of the inner and outer $R$ channels in each octant.
\bibitem{88mcnote}
This is in contrast to the analysis of data from the commissioning run where we
  used just empirical parameterizations of the electromagnetic showers in
  simulating the EMLAC response.
\end{thebibliography}
\bibliographystyle{prsty}

%
%
\widetext
\newpage
\noindent
\newpage
\newpage
\newpage
\newpage
\newpage
\newpage
\newpage
\newpage
\newpage
\newpage
\newpage
\newpage
\newpage
\newpage
\newpage
\newpage
\newpage
\newpage
\newpage
\newpage
\newpage
\newpage
\newpage
\newpage
%
%
\newpage
\narrowtext
\begin{table}
\caption{
Compilation of systematic uncertainties for inclusive \piz\ and $\eta$ 
cross sections. The ranges correspond to variations between low-\pt\ 
to high-\pt\ values.}
\begin{tabular}{ccc}
Source\\ of Uncertainty &  \piz\ production & $\eta$ production \\ 
\tableline
Luminosity         	& $8\%$   & $8\%$   \\
Energy Scale 		& $5\% \rightarrow 9\%$ & $5\% \rightarrow 9\%$ \\
Beam Contamination (530 GeV/c beam) & $0\% \rightarrow 7\%$ & $0\% \rightarrow 7\%$ \\
Reconstruction Efficiency 	& $5\%$   & $6\%$  \\
Trigger 		& $2\% \rightarrow 0\%$ & $5\% \rightarrow 0\%$ \\
Beam Halo Muon Rejection	       	& $1.5\%$ & $1\%$ \\
Geometrical Acceptance 	& $1\%$   & $1\%$   \\
Photon Conversions  	& $1\%$   & $1\%$   \\
Vertex Finding      	& $1\%$   & $1\%$   \\
Background Subtraction 	& $2\% \rightarrow 0\%$   & $2\%$   \\
\tableline
Total			& $11\% \rightarrow 15\%$ & $13\% \rightarrow 15\%$ \\
\end{tabular}
\label{syserrs}
\end{table}
\newpage
\widetext
\begin{table}
\caption{
Invariant differential cross sections $\left( Ed^3 \sigma/dp^3 \right)$ 
per nucleon for the inclusive reaction \pBe\ $\rightarrow$ \piz X at 530 and 
800~GeV/$c$, averaged over the rapidity intervals $-0.75 \le \ycm \le 0.75$ 
and $-1.0 \le \ycm \le 0.5$, respectively.}
\squeezetable
\begin{tabular}{r@{ -- }l
r@{${}\pm{}$}l
r@{${}\pm{}$}l}
\multicolumn{2}{c}{$p_T$}  &\multicolumn{2}{c}{${\mit p}$Be @   $530~GeV/c$}  &\multicolumn{2}{c}{${\mit p}$Be @   $800~GeV/c$} \\
\multicolumn{2}{c}{$(GeV/c)$}  &\multicolumn{2}{c}{$(\mu{b}/(GeV/c)^2)$}  &\multicolumn{2}{c}{$(\mu{b}/(GeV/c)^2)$} \\
 \tableline
  1.00 & 1.20 & 
467 & 29${}\pm{}$49 & 548 & 49${}\pm{}$61 \\
  1.20 & 1.40 & 
199 & 14${}\pm{}$21 & 267 & 23${}\pm{}$30 \\
  1.40 & 1.60 & 
68.4 & 7.3${}\pm{}$7.2 & 107 & 11${}\pm{}$12 \\
  1.60 & 1.80 & 
26.7 & 3.8${}\pm{}$2.8 & 49.3 & 5.7${}\pm{}$5.5 \\
  1.80 & 2.00 & 
15.1 & 2.1${}\pm{}$1.6 & 21.5 & 3.2${}\pm{}$2.4 \\
  2.00 & 2.20 & 
6.7 & 1.3${}\pm{}$0.71 & 7.3 & 1.7${}\pm{}$0.82 \\
\end{tabular}
\label{pi0_table_be}
\end{table}
\newpage
\widetext
\begin{table}
\caption{
Invariant differential cross sections $\left( Ed^3 \sigma/dp^3 \right)$
for the inclusive reaction \pp\ $\rightarrow$ \piz X at 530 and 800~GeV/$c$,
averaged over the rapidity intervals $-0.75 \le \ycm \le 0.75$ and
$-1.0 \le \ycm \le 0.5$, respectively.}
\squeezetable
\begin{tabular}{r@{ -- }l
r@{${}\pm{}$}l
r@{${}\pm{}$}l}
\multicolumn{2}{c}{$p_T$}  &\multicolumn{2}{c}{${\mit pp}$ @    $530~GeV/c$}  &\multicolumn{2}{c}{${\mit pp}$ @    $800~GeV/c$} \\
\multicolumn{2}{c}{$(GeV/c)$}  &\multicolumn{2}{c}{$(\mu{b}/(GeV/c)^2)$}  &\multicolumn{2}{c}{$(\mu{b}/(GeV/c)^2)$} \\
 \tableline
  1.00 & 1.40 & 
284 & 44${}\pm{}$30 & 531 & 72${}\pm{}$59 \\
  1.40 & 1.80 & 
63 & 10${}\pm{}$6.7 & 113 & 17${}\pm{}$13 \\
  1.80 & 2.20 & 
13.5 & 3.6${}\pm{}$1.4 & 17.3 & 5.2${}\pm{}$1.9 \\
  2.20 & 2.40 & 
3.36 & 0.81${}\pm{}$0.36 & 3.25 & 0.43${}\pm{}$0.36 \\
  2.40 & 2.60 & 
1.15 & 0.14${}\pm{}$0.12 & 1.33 & 0.22${}\pm{}$0.15 \\
 \tableline
\multicolumn{2}{c}{ }  &\multicolumn{2}{c}{$(nb/(GeV/c)^2)$}  &\multicolumn{2}{c}{$(nb/(GeV/c)^2)$} \\
\end{tabular}
\label{pi0_table_p}
\end{table}
\widetext
\newpage
\begin{table}
\caption{
The averaged invariant differential cross section per nucleon
$\left( Ed^3 \sigma/dp^3 \right)$ as a function of rapidity and \pt\ 
for the inclusive reaction \pBe~$\rightarrow$~\piz X at 530~GeV/$c$.}
\squeezetable
\begin{tabular}{r@{ -- }l
r@{ }l
r@{ }l
r@{ }l
r@{ }l}
\multicolumn{2}{c}{$y_{cm}$}  &\multicolumn{8}{c}{$p_T$ (GeV/c)} \\
\multicolumn{2}{c}{ } &\multicolumn{2}{c}{1.00{ -- }1.50 } &\multicolumn{2}{c}{1.50{ -- }2.00 } &\multicolumn{2}{c}{2.00{ -- }2.50 } &\multicolumn{2}{c}{2.50{ -- }3.00 } \\
\multicolumn{2}{c}{ } &\multicolumn{2}{c}{$(\mu{b}/(GeV/c)^2)$} &\multicolumn{2}{c}{$(\mu{b}/(GeV/c)^2)$} &\multicolumn{2}{c}{$(\mu{b}/(GeV/c)^2)$} &\multicolumn{2}{c}{$(nb/(GeV/c)^2)$} \\
 \tableline
  $-0.750$ & $-0.625$ & 
\vctr{388${\,}\pm$} & \vctr{43${}\pm{}$41} & \vctr{27.2${\,}\pm$} & \vctr{7.9${}\pm{}$2.9} & \vctr{6.5${\,}\pm$} & \vctr{1.6${}\pm{}$0.7} & {820${\,}\pm$} & {180${}\pm{}$90} \\
  $-0.625$ & $-0.500$ & 
{} & {} & {} & {} & {} & {} & {409${\,}\pm$} & {86${}\pm{}$44} \\
  $-0.500$ & $-0.375$ & 
\vctr{313${\,}\pm$} & \vctr{40${}\pm{}$33} & \vctr{25.5${\,}\pm$} & \vctr{6.0${}\pm{}$2.7} & \vctr{2.9${\,}\pm$} & \vctr{1.2${}\pm{}$0.3} & {496${\,}\pm$} & {23${}\pm{}$53} \\
  $-0.375$ & $-0.250$ & 
{} & {} & {} & {} & {} & {} & {494${\,}\pm$} & {15${}\pm{}$53} \\
  $-0.250$ & $-0.125$ & 
\vctr{324${\,}\pm$} & \vctr{33${}\pm{}$34} & \vctr{29.7${\,}\pm$} & \vctr{6.2${}\pm{}$3.1} & \vctr{4.1${\,}\pm$} & \vctr{1.2${}\pm{}$0.4} & {513${\,}\pm$} & {12${}\pm{}$55} \\
\end{tabular}
\label{pi0_table_be_530_rap}
\end{table}
\newpage
\widetext
\begin{table}
\caption{
The averaged invariant differential cross section 
$\left( Ed^3 \sigma/dp^3 \right)$ as a function of rapidity and \pt\ 
for the inclusive reaction \pp~$\rightarrow$ \piz X at 530~GeV/$c$.}
\squeezetable
\begin{tabular}{r@{ -- }l
r@{ }l
r@{ }l
r@{ }l
r@{ }l}
\multicolumn{2}{c}{$y_{cm}$}  &\multicolumn{8}{c}{$p_T$ (GeV/c)} \\
\multicolumn{2}{c}{ } &\multicolumn{2}{c}{1.00{ -- }2.50 } &\multicolumn{2}{c}{2.50{ -- }3.00 } &\multicolumn{2}{c}{3.00{ -- }3.50 } &\multicolumn{2}{c}{3.50{ -- }4.00 } \\
\multicolumn{2}{c}{ } &\multicolumn{2}{c}{$(\mu{b}/(GeV/c)^2)$} &\multicolumn{2}{c}{$(nb/(GeV/c)^2)$} &\multicolumn{2}{c}{$(nb/(GeV/c)^2)$} &\multicolumn{2}{c}{$(nb/(GeV/c)^2)$} \\
 \tableline
  $-0.750$ & $-0.625$ & 
\vctr{147${\,}\pm$} & \vctr{15${}\pm{}$16} & {540${\,}\pm$} & {180${}\pm{}$60} & {68${\,}\pm$} & {10${}\pm{}$7} & {8.7${\,}\pm$} & {1.4${}\pm{}$0.9} \\
  $-0.625$ & $-0.500$ & 
{} & {} & {950${\,}\pm$} & {330${}\pm{}$100} & {66${\,}\pm$} & {15${}\pm{}$7} & {10.0${\,}\pm$} & {1.4${}\pm{}$1.1} \\
  $-0.500$ & $-0.375$ & 
\vctr{119${\,}\pm$} & \vctr{15${}\pm{}$13} & {365${\,}\pm$} & {40${}\pm{}$39} & {61${\,}\pm$} & {10${}\pm{}$6} & {12.3${\,}\pm$} & {2.7${}\pm{}$1.3} \\
  $-0.375$ & $-0.250$ & 
{} & {} & {440${\,}\pm$} & {39${}\pm{}$47} & {65.5${\,}\pm$} & {7.2${}\pm{}$7.0} & {14.9${\,}\pm$} & {2.3${}\pm{}$1.6} \\
  $-0.250$ & $-0.125$ & 
\vctr{123${\,}\pm$} & \vctr{12${}\pm{}$13} & {458${\,}\pm$} & {26${}\pm{}$49} & {76.7${\,}\pm$} & {6.9${}\pm{}$8.2} & {15.4${\,}\pm$} & {2.4${}\pm{}$1.6} \\
\end{tabular}
\label{pi0_table_p_530_rap}
\end{table}
\begin{table}
\caption{
The averaged invariant differential cross sections per nucleon
$\left( Ed^3 \sigma/dp^3 \right)$ as a function of rapidity and \pt\ 
for the inclusive reaction \pBe~$\rightarrow$ \piz X at 800~GeV/$c$.}
\squeezetable
\begin{tabular}{r@{ -- }l
r@{ }l
r@{ }l
r@{ }l
r@{ }l}
\multicolumn{2}{c}{$y_{cm}$}  &\multicolumn{8}{c}{$p_T$ (GeV/c)} \\
\multicolumn{2}{c}{ } &\multicolumn{2}{c}{1.00{ -- }1.50 } &\multicolumn{2}{c}{1.50{ -- }2.00 } &\multicolumn{2}{c}{2.00{ -- }2.50 } &\multicolumn{2}{c}{2.50{ -- }3.00 } \\
\multicolumn{2}{c}{ } &\multicolumn{2}{c}{$(\mu{b}/(GeV/c)^2)$} &\multicolumn{2}{c}{$(\mu{b}/(GeV/c)^2)$} &\multicolumn{2}{c}{$(\mu{b}/(GeV/c)^2)$} &\multicolumn{2}{c}{$(nb/(GeV/c)^2)$} \\
 \tableline
  $-1.00$  & $-0.875$ & 
\vctr{444${\,}\pm$} & \vctr{71${}\pm{}$49} & \vctr{36${\,}\pm$} & \vctr{11${}\pm{}$4} & \vctr{6.5${\,}\pm$} & \vctr{1.9${}\pm{}$0.7} & {477${\,}\pm$} & {87${}\pm{}$53} \\
  $-0.875$ & $-0.750$ & 
{} & {} & {} & {} & {} & {} & {840${\,}\pm$} & {180${}\pm{}$90} \\
  $-0.750$ & $-0.625$ & 
\vctr{297${\,}\pm$} & \vctr{65${}\pm{}$33} & \vctr{58.8${\,}\pm$} & \vctr{9.0${}\pm{}$6.5} & \vctr{5.9${\,}\pm$} & \vctr{1.8${}\pm{}$0.7} & {615${\,}\pm$} & {55${}\pm{}$68} \\
  $-0.625$ & $-0.500$ & 
{} & {} & {} & {} & {} & {} & {686${\,}\pm$} & {50${}\pm{}$76} \\
  $-0.500$ & $-0.375$ & 
\vctr{372${\,}\pm$} & \vctr{59${}\pm{}$41} & \vctr{57.1${\,}\pm$} & \vctr{9.4${}\pm{}$6.3} & \vctr{1.3${\,}\pm$} & \vctr{1.8${}\pm{}$0.1} & {610${\,}\pm$} & {50${}\pm{}$68} \\
\end{tabular}
\label{pi0_table_be_800_rap}
\end{table}
\begin{table}
\caption{
The averaged invariant differential cross sections
$\left( Ed^3 \sigma/dp^3 \right)$ as a function of rapidity and \pt\ 
for the inclusive reaction \pp~$\rightarrow$ \piz X at 800~GeV/$c$.}
\squeezetable
\begin{tabular}{r@{ -- }l
r@{ }l
r@{ }l
r@{ }l
r@{ }l}
\multicolumn{2}{c}{$y_{cm}$}  &\multicolumn{8}{c}{$p_T$ (GeV/c)} \\
\multicolumn{2}{c}{ } &\multicolumn{2}{c}{1.00{ -- }2.50 } &\multicolumn{2}{c}{2.50{ -- }3.00 } &\multicolumn{2}{c}{3.00{ -- }3.50 } &\multicolumn{2}{c}{3.50{ -- }4.00 } \\
\multicolumn{2}{c}{ } &\multicolumn{2}{c}{$(\mu{b}/(GeV/c)^2)$} &\multicolumn{2}{c}{$(nb/(GeV/c)^2)$} &\multicolumn{2}{c}{$(nb/(GeV/c)^2)$} &\multicolumn{2}{c}{$(nb/(GeV/c)^2)$} \\
 \tableline
  $-1.00$  & $-0.875$ & 
\vctr{166${\,}\pm$} & \vctr{26${}\pm{}$18} & {980${\,}\pm$} & {580${}\pm{}$110} & {83${\,}\pm$} & {16${}\pm{}$9} & {19.2${\,}\pm$} & {7.4${}\pm{}$2.1} \\
  $-0.875$ & $-0.750$ & 
{} & {} & {410${\,}\pm$} & {110${}\pm{}$50} & {108${\,}\pm$} & {14${}\pm{}$12} & {18.7${\,}\pm$} & {4.1${}\pm{}$2.1} \\
  $-0.750$ & $-0.625$ & 
\vctr{120${\,}\pm$} & \vctr{24${}\pm{}$13} & {420${\,}\pm$} & {63${}\pm{}$47} & {101${\,}\pm$} & {14${}\pm{}$11} & {36.9${\,}\pm$} & {6.9${}\pm{}$4.1} \\
  $-0.625$ & $-0.500$ & 
{} & {} & {570${\,}\pm$} & {100${}\pm{}$60} & {111${\,}\pm$} & {12${}\pm{}$12} & {18.0${\,}\pm$} & {3.6${}\pm{}$2.0} \\
  $-0.500$ & $-0.375$ & 
\vctr{146${\,}\pm$} & \vctr{21${}\pm{}$16} & {560${\,}\pm$} & {110${}\pm{}$60} & {144${\,}\pm$} & {14${}\pm{}$16} & {26.6${\,}\pm$} & {5.5${}\pm{}$2.9} \\
\end{tabular}
\label{pi0_table_p_800_rap}
\end{table}
\widetext
\begin{table}
\caption{
Invariant differential cross sections $\left( Ed^3 \sigma/dp^3 \right)$
per nucleon for the inclusive reaction \pBe~$\rightarrow \eta $X at 530 and
800~GeV/$c$, averaged over the rapidity intervals $-0.75 \le \ycm \le 0.75$ 
and $-1.0 \le \ycm \le 0.5$, respectively.}
\squeezetable
\begin{tabular}{r@{ -- }l
r@{${}\pm{}$}l
r@{${}\pm{}$}l}
\multicolumn{2}{c}{$p_T$}  &\multicolumn{2}{c}{${\mit p}$Be @   $530~GeV/c$}  &\multicolumn{2}{c}{${\mit p}$Be @   $800~GeV/c$} \\
\multicolumn{2}{c}{$(GeV/c)$}  &\multicolumn{2}{c}{$(nb/(GeV/c)^2)$}  &\multicolumn{2}{c}{$(nb/(GeV/c)^2)$} \\
 \tableline
  3.00 & 3.20 & 
49.2 & 9.0${}\pm{}$6.0 & 88 & 18${}\pm{}$11 \\
  3.20 & 3.40 & 
24.4 & 4.5${}\pm{}$2.9 & 34.7 & 8.5${}\pm{}$4.3 \\
  3.40 & 3.60 & 
16.3 & 2.5${}\pm{}$1.9 & 19.5 & 5.1${}\pm{}$2.4 \\
  3.60 & 3.80 & 
8.5 & 1.4${}\pm{}$0.98 & 8.1 & 3.0${}\pm{}$0.96 \\
  3.80 & 4.00 & 
3.67 & 0.25${}\pm{}$0.42 & 7.34 & 0.77${}\pm{}$0.86 \\
  4.00 & 4.20 & 
2.00 & 0.13${}\pm{}$0.23 & 3.89 & 0.33${}\pm{}$0.45 \\
  4.20 & 4.40 & 
1.242 & 0.078${}\pm{}$0.14 & 2.18 & 0.19${}\pm{}$0.25 \\
  4.40 & 4.60 & 
0.587 & 0.043${}\pm{}$0.067 & 1.20 & 0.12${}\pm{}$0.14 \\
  4.60 & 4.80 & 
0.409 & 0.024${}\pm{}$0.047 & 0.805 & 0.073${}\pm{}$0.095 \\
  4.80 & 5.00 & 
0.244 & 0.017${}\pm{}$0.028 & 0.502 & 0.048${}\pm{}$0.059 \\
 \tableline
\multicolumn{2}{c}{ }  &\multicolumn{2}{c}{$(pb/(GeV/c)^2)$}  &\multicolumn{2}{c}{$(pb/(GeV/c)^2)$} \\
 \tableline
  5.00 & 5.25 & 
132.5 & 9.2${}\pm{}$15 & 341 & 27${}\pm{}$40 \\
  5.25 & 5.50 & 
70.2 & 5.6${}\pm{}$8.2 & 152 & 15${}\pm{}$18 \\
  5.50 & 5.75 & 
36.3 & 4.4${}\pm{}$4.3 & 107 & 12${}\pm{}$13 \\
  5.75 & 6.00 & 
17.2 & 2.2${}\pm{}$2.0 & 58.0 & 8.0${}\pm{}$7.0 \\
  6.00 & 6.50 & 
7.27 & 0.98${}\pm{}$0.87 & 26.3 & 3.6${}\pm{}$3.2 \\
  6.50 & 7.00 & 
2.63 & 0.43${}\pm{}$0.32 & 10.4 & 1.8${}\pm{}$1.3 \\
  7.00 & 8.00 & 
0.23 & 0.13${}\pm{}$0.028 & 3.02 & 0.64${}\pm{}$0.38 \\
  8.00 & 9.00 & 
0.035 & 0.046${}\pm{}$0.005 & 1.33 & 0.43${}\pm{}$0.17 \\
   9.00 & 10.00 & 
\multicolumn{2}{c}{---} & 0.11 & 0.24${}\pm{}$0.015 \\
  10.00 & 12.00 & 
\multicolumn{2}{c}{---} & 0.073 & 0.050${}\pm{}$0.010 \\
 \end{tabular}
\label{eta_table_be}
\end{table}
\widetext
\begin{table}
\caption{
Invariant differential cross sections $\left( Ed^3 \sigma/dp^3 \right)$
for the inclusive reaction \pp~$\rightarrow \eta $X at 530 and 800~GeV/$c$, 
averaged over the rapidity intervals $-0.75 \le \ycm \le 0.75$ and
$-1.0 \le \ycm \le 0.5$, respectively.}
\squeezetable
\begin{tabular}{r@{ -- }l
r@{${}\pm{}$}l
r@{${}\pm{}$}l}
\multicolumn{2}{c}{$p_T$}  &\multicolumn{2}{c}{${\mit pp}$ @    $530~GeV/c$}  &\multicolumn{2}{c}{${\mit pp}$ @    $800~GeV/c$} \\
\multicolumn{2}{c}{$(GeV/c)$}  &\multicolumn{2}{c}{$(nb/(GeV/c)^2)$}  &\multicolumn{2}{c}{$(nb/(GeV/c)^2)$} \\
 \tableline
  3.00 & 3.50 & 
22.6 & 9.5${}\pm{}$2.7 & 51 & 17${}\pm{}$6.3 \\
  3.50 & 4.00 & 
7.6 & 1.9${}\pm{}$0.87 & 11.4 & 3.7${}\pm{}$1.4 \\
  4.00 & 4.50 & 
1.19 & 0.14${}\pm{}$0.13 & 2.13 & 0.34${}\pm{}$0.25 \\
  4.50 & 5.00 & 
0.310 & 0.035${}\pm{}$0.035 & 0.678 & 0.091${}\pm{}$0.080 \\
 \tableline
\multicolumn{2}{c}{ }  &\multicolumn{2}{c}{$(pb/(GeV/c)^2)$}  &\multicolumn{2}{c}{$(pb/(GeV/c)^2)$} \\
 \tableline
  5.00 & 5.50 & 
77 & 12${}\pm{}$9.0 & 220 & 36${}\pm{}$26 \\
  5.50 & 6.00 & 
23.7 & 5.0${}\pm{}$2.8 & 108 & 16${}\pm{}$13 \\
  6.00 & 7.00 & 
2.3 & 1.2${}\pm{}$0.3 & 20.6 & 4.8${}\pm{}$2.5 \\
  7.00 & 8.00 & 
0.41 & 0.31${}\pm{}$0.05 & 4.4 & 1.8${}\pm{}$0.6 \\
   8.00 & 10.00 & 
\multicolumn{2}{c}{---} & 1.05 & 0.70${}\pm{}$0.14 \\
 \end{tabular}
\label{eta_table_p}
\end{table}
\widetext
\begin{table}
\caption{
The averaged invariant differential cross sections per nucleon
$\left( Ed^3 \sigma/dp^3 \right)$ as a function of rapidity and \pt\ 
for the inclusive reaction \pBe~$\rightarrow \eta $X at 530~GeV/$c$.
The units are $\rm{pb}/(GeV/c)^2$.}
\squeezetable
\begin{tabular}{r@{ -- }l
r@{ }l
r@{ }l
r@{ }l
r@{ }l}
\multicolumn{2}{c}{$y_{cm}$}  &\multicolumn{8}{c}{$p_T$ (GeV/c)} \\
\multicolumn{2}{c}{ } &\multicolumn{2}{c}{3.00{ -- }4.00 } &\multicolumn{2}{c}{4.00{ -- }4.50 } &\multicolumn{2}{c}{4.50{ -- }5.00 } &\multicolumn{2}{c}{5.00{ -- }5.50 } \\
 \tableline
  $-0.750$ & $-0.625$ & 
\vctr{14600${\,}\pm$} & \vctr{9000${}\pm{}$1600} & {1050${\,}\pm$} & {310${}\pm{}$110} & {247${\,}\pm$} & {62${}\pm{}$27} & {66${\,}\pm$} & {25${}\pm{}$7} \\
  $-0.625$ & $-0.500$ & 
{} & {} & {950${\,}\pm$} & {320${}\pm{}$100} & {318${\,}\pm$} & {88${}\pm{}$34} & {96${\,}\pm$} & {20${}\pm{}$10} \\
  $-0.500$ & $-0.375$ & 
\vctr{24100${\,}\pm$} & \vctr{4900${}\pm{}$2600} & {1420${\,}\pm$} & {280${}\pm{}$150} & {374${\,}\pm$} & {68${}\pm{}$40} & {100${\,}\pm$} & {26${}\pm{}$11} \\
  $-0.375$ & $-0.250$ & 
{} & {} & {1730${\,}\pm$} & {290${}\pm{}$190} & {431${\,}\pm$} & {60${}\pm{}$47} & {111${\,}\pm$} & {18${}\pm{}$12} \\
  $-0.250$ & $-0.125$ & 
\vctr{21200${\,}\pm$} & \vctr{3800${}\pm{}$2300} & {1930${\,}\pm$} & {250${}\pm{}$210} & {486${\,}\pm$} & {57${}\pm{}$52} & {134${\,}\pm$} & {19${}\pm{}$15} \\
  $-0.125$ & $0.000$  & 
{} & {} & {1750${\,}\pm$} & {200${}\pm{}$190} & {514${\,}\pm$} & {56${}\pm{}$56} & {123${\,}\pm$} & {17${}\pm{}$13} \\
\end{tabular}
\label{eta_table_be_530_rap}
\end{table}
\widetext
\begin{table}
\caption{
The averaged invariant differential cross sections
$\left( Ed^3 \sigma/dp^3 \right)$ as a function of rapidity and \pt\ 
for the inclusive reaction \pp~$\rightarrow \eta $X at 530~GeV/$c$.
The units are $\rm{pb}/(GeV/c)^2$.}
\squeezetable
\begin{tabular}{r@{ -- }l
r@{ }l
r@{ }l
r@{ }l}
\multicolumn{2}{c}{$y_{cm}$}      &\multicolumn{6}{c}{$p_T$ (GeV/c)} \\
\multicolumn{2}{c}{ } &\multicolumn{2}{c}{4.00{ -- }5.00 } &\multicolumn{2}{c}{5.00{ -- }6.00 } &\multicolumn{2}{c}{6.00{ -- }8.00 } \\
 \tableline
  $-0.75$  & $-0.50$  & 
{520${\,}\pm$} & {250${}\pm{}$60} & {29${\,}\pm$} & {19${}\pm{}$3} & \vctr{1.6${\,}\pm$} & \vctr{1.0${}\pm{}$0.2} \\
  $-0.50$  & $-0.25$  & 
{700${\,}\pm$} & {230${}\pm{}$80} & {53${\,}\pm$} & {16${}\pm{}$6} & {} & {} \\
  $-0.25$  & $0.00$   & 
{1150${\,}\pm$} & {190${}\pm{}$120} & {57${\,}\pm$} & {16${}\pm{}$6} & \vctr{1.1${\,}\pm$} & \vctr{1.0${}\pm{}$0.1} \\
  $0.00$   & $0.25$   & 
{740${\,}\pm$} & {150${}\pm{}$80} & {70${\,}\pm$} & {18${}\pm{}$8} & {} & {} \\
  $0.25$   & $0.50$   & 
{800${\,}\pm$} & {110${}\pm{}$90} & {46${\,}\pm$} & {13${}\pm{}$5} & \vctr{1.3${\,}\pm$} & \vctr{1.2${}\pm{}$0.1} \\
  $0.50$   & $0.75$   & 
{573${\,}\pm$} & {73${}\pm{}$61} & {48${\,}\pm$} & {13${}\pm{}$5} & {} & {} \\
 \end{tabular}
\label{eta_table_p_530_rap}
\end{table}
\widetext
\begin{table}
\caption{
The averaged invariant differential cross sections per nucleon
$\left( Ed^3 \sigma/dp^3 \right)$ as a function of rapidity and \pt\ 
for the inclusive reaction \pBe~$\rightarrow \eta $X at 800~GeV/$c$.
The units are $\rm{pb}/(GeV/c)^2$.}
\squeezetable
\begin{tabular}{r@{ -- }l
r@{ }l
r@{ }l
r@{ }l
r@{ }l}
\multicolumn{2}{c}{$y_{cm}$}  &\multicolumn{8}{c}{$p_T$ (GeV/c)} \\
\multicolumn{2}{c}{ } &\multicolumn{2}{c}{3.00{ -- }4.00 } &\multicolumn{2}{c}{4.00{ -- }4.50 } &\multicolumn{2}{c}{4.50{ -- }5.00 } &\multicolumn{2}{c}{5.00{ -- }5.50 } \\
 \tableline
  $-1.00$  & $-0.875$ & 
\vctr{13000${\,}\pm$} & \vctr{15000${}\pm{}$1000} & {920${\,}\pm$} & {580${}\pm{}$100} & \multicolumn{2}{c}{---} & {116${\,}\pm$} & {47${}\pm{}$13} \\
  $-0.875$ & $-0.750$ & 
{} & {} & {1360${\,}\pm$} & {620${}\pm{}$150} & {500${\,}\pm$} & {240${}\pm{}$60} & {87${\,}\pm$} & {52${}\pm{}$10} \\
  $-0.750$ & $-0.625$ & 
\vctr{46000${\,}\pm$} & \vctr{11000${}\pm{}$5000} & {1980${\,}\pm$} & {630${}\pm{}$220} & {560${\,}\pm$} & {130${}\pm{}$60} & {155${\,}\pm$} & {43${}\pm{}$17} \\
  $-0.625$ & $-0.500$ & 
{} & {} & {1270${\,}\pm$} & {580${}\pm{}$140} & {540${\,}\pm$} & {160${}\pm{}$60} & {259${\,}\pm$} & {48${}\pm{}$29} \\
  $-0.500$ & $-0.375$ & 
\vctr{17300${\,}\pm$} & \vctr{8100${}\pm{}$1900} & {2330${\,}\pm$} & {550${}\pm{}$260} & {640${\,}\pm$} & {120${}\pm{}$70} & {299${\,}\pm$} & {49${}\pm{}$34} \\
  $-0.375$ & $-0.250$ & 
{} & {} & {3840${\,}\pm$} & {530${}\pm{}$420} & {710${\,}\pm$} & {150${}\pm{}$80} & {253${\,}\pm$} & {56${}\pm{}$29} \\
\end{tabular}
\label{eta_table_be_800_rap}
\end{table}
\widetext
\begin{table}
\caption{The averaged invariant differential cross sections
$\left( Ed^3 \sigma/dp^3 \right)$ as a function of rapidity and \pt\ 
for the inclusive reaction \pp~$\rightarrow \eta $X at 800~GeV/$c$.
The units are $\rm{pb}/(GeV/c)^2$.}
\squeezetable
\begin{tabular}{r@{ -- }l
r@{ }l
r@{ }l
r@{ }l}
\multicolumn{2}{c}{$y_{cm}$}      &\multicolumn{6}{c}{$p_T$ (GeV/c)} \\
\multicolumn{2}{c}{ } &\multicolumn{2}{c}{4.00{ -- }5.00 } &\multicolumn{2}{c}{5.00{ -- }6.00 } &\multicolumn{2}{c}{6.00{ -- }8.00 } \\
 \tableline
  $-1.0$   & $-0.75$  & 
{1060${\,}\pm$} & {430${}\pm{}$120} & {73${\,}\pm$} & {50${}\pm{}$8} & {6.9${\,}\pm$} & {3.6${}\pm{}$0.8} \\
  $-0.75$  & $-0.50$  & 
{360${\,}\pm$} & {520${}\pm{}$40} & {97${\,}\pm$} & {49${}\pm{}$11} & {4.9${\,}\pm$} & {7.1${}\pm{}$0.6} \\
  $-0.50$  & $-0.25$  & 
{1930${\,}\pm$} & {420${}\pm{}$210} & {194${\,}\pm$} & {40${}\pm{}$22} & {14.4${\,}\pm$} & {7.6${}\pm{}$1.7} \\
  $-0.25$  & $0.00$   & 
{1880${\,}\pm$} & {440${}\pm{}$210} & {256${\,}\pm$} & {47${}\pm{}$29} & {22.3${\,}\pm$} & {6.0${}\pm{}$2.6} \\
  $0.00$   & $0.25$   & 
{1980${\,}\pm$} & {470${}\pm{}$220} & {167${\,}\pm$} & {55${}\pm{}$19} & {16.9${\,}\pm$} & {7.6${}\pm{}$2.0} \\
  $0.25$   & $0.50$   & 
{1230${\,}\pm$} & {250${}\pm{}$140} & {198${\,}\pm$} & {50${}\pm{}$22} & {9.6${\,}\pm$} & {5.0${}\pm{}$1.1} \\
 \end{tabular}
\label{eta_table_p_800_rap}
\end{table}
\end{document}